\documentclass[submission,Phys]{SciPost}

\usepackage{amsmath}
\usepackage{amssymb}
\usepackage{graphicx}
\usepackage{dcolumn}
\usepackage{bm}
\usepackage{xcolor}
\usepackage{hyperref}
\usepackage{xcolor}

\hypersetup{
    colorlinks,
    linkcolor={red!30!black},
    citecolor={blue!50!black},
    urlcolor={blue!80!black}
}

\DeclareSymbolFont{usualmathcal}{OMS}{cmsy}{m}{n}
\DeclareSymbolFontAlphabet{\mathcal}{usualmathcal}

\begin{document}

\begin{center}
{\Large \textbf{Emptiness Formation in Polytropic Quantum Liquids}}
\end{center}

\begin{center}
Hsiu-Chung Yeh\textsuperscript{1*}, 
Dimitri M. Gangardt\textsuperscript{2},
Alex Kamenev \textsuperscript{1,3}
\end{center}

\begin{center}
{\bf 1} School of Physics and Astronomy, University of Minnesota, \\  Minneapolis, Minnesota 55455, USA
\\
{\bf 2} School of Physics and Astronomy, University of Birmingham,
  Edgbaston, \\ Birmingham, B15 2TT, United Kingdom
 \\
 {\bf 3} William I. Fine Theoretical Physics Institute, University of
 Minnesota, Minneapolis,  Minnesota 55455, USA
 \\
 \bigskip
* yehxx129@umn.edu
\bigskip
\end{center}

\begin{center}
\today
\end{center}

\section*{Abstract}
{\bf
  We study large deviations in interacting quantum liquids with the polytropic
  equation of state $P(\rho)\sim \rho^\gamma$, where $\rho$ is density and $P$
  is pressure. By solving hydrodynamic equations in imaginary time we evaluate
  the instanton action and calculate the emptiness formation probability
  (EFP), the probability that no particle resides in a macroscopic interval of
  a given size.  Analytic solutions are found for a certain infinite sequence
  of rational polytropic indexes $\gamma$ and the result can be analytically
  continued to any value of $\gamma\ge 1$. Our findings agree with (and
  significantly expand on) previously known analytical and numerical results
  for EFP in quantum liquids. We also discuss interesting universal spacetime
  features of the instanton solution.
}



\vspace{10pt}
\noindent\rule{\textwidth}{1pt}
\tableofcontents\thispagestyle{fancy}
\noindent\rule{\textwidth}{1pt}
\vspace{10pt}

\section{Introduction}

Large deviations statistics in many-body systems has been a subject of the rapidly growing research efforts  \cite{del2011long,pons2012fidelity,del2016exact,arzamasovs2019full} due to recent precision measurements of particle number fluctuations in ultra cold quantum gases \cite{esteve2006observations,armijo2010probing,jacqmin2011sub}. Emptiness formation probability (EFP) is perhaps the most iconic and widely studied example of such large deviation. It is accessible through the Bethe ansatz  \cite{bethe1931theorie} in certain integrable models \cite{korepin1994correlation,korepin1997quantum,de2001six,boos2003emptiness} and serves as the litmus test for validity of approximate non-perturbative techniques, such as the instanton calculus \cite{kleinert2009path}. 

The EFP, ${\cal P}_\mathrm{EFP}(R)$, is the probability that no particles are found inside the space interval $[-R,R]$ in the ground state of e.g. a one-dimensional (1D) many-body system
\begin{align}
\label{eq:EFP}
{\cal P}_\mathrm{EFP}(R) =  \prod_{i=1}^N \int_{|x_i| \geq R} dx_i\ |\Psi_\mathrm{GS}(x_1,x_2,...,x_N)|^2.
\end{align}
Here $\Psi_\mathrm{GS}(x_1,x_2,...,x_N)$ is the normalized ground state wave
function of the N-particle system. Even in cases where $\Psi_\mathrm{GS}$ is
known exactly (e.g. for free fermions, or through the Bethe Ansatz), it is
still a formidable task to perform the multiple integrals over the restricted
interval. The first discussion of such problem goes back to the random matrix
theory (RMT) \cite{mehta2004random}, where the probability that no eigenvalues
are located within a certain energy interval was studied for different
ensembles \cite{dyson1962statisticalII,dyson1962statisticalIII}.

Going beyond free fermions and random matrices, integrable spin-$1/2$ chains
are probably the most studied systems in the  context of EFP. The later  is
defined as the probability of measuring $l$ aligned ``up'' spins in the ground
state. Via the Jordan-Wigner transformation the problem becomes equivalent to
the absence of quasiparticles on $l$ consecutive sites
\cite{shiroishi2001emptiness} and EFP was found in terms of
Fredholm determinants
\cite{korepin1994correlation,kitanine2000correlation,shiroishi2001emptiness,kitanine2002emptiness}.
The closed analytic expressions for EFP are available only in a few isolated
cases in the parameter space \cite{shiroishi2001emptiness,Kitanine_2002}.

With few exceptions \cite{bastianello2018exact,arzamasovs2019full}
most studies have been focused on the asymptotic regime of large $R$, where
EFP is exponentially small. This makes EFP suitable for
semiclassical instanton approach, where
$-\ln {\cal P}_\mathrm{EFP}(R)$ is given  by classical action evaluated along a
stationary trajectory of the imaginary time Euler-Lagrange equations
\cite{kleinert2009path}. Such trajectory is specified by imposing boundary
conditions both in the distant ``past'' and ``future'', when the system
  is undisturbed, and at the observation time,  when the rare fluctuation develops.
Similar setup also shows up in
studies of rare events in classical stochastic systems
\cite{dykman1994large,elgart2006classification,krapivsky2012void}.  Even for
classically integrable equations (such as eg. via inverse scattering
technique) these problems are notoriously difficult to handle (for a very
recent progress in this direction see Refs.~
\cite{krajenbrink2021PRL,krajenbrink2021inverse}). Although reasonably
effective numerical methods has been developed
\cite{chernykh2001large,elgart2004rare,janas2016dynamical}, their applications
still require significant time and computer resources.

In this paper, we focus on 1D polytropic liquid which is characterized by
equation of state: $P(\rho) \sim \rho^\gamma$, where $P(\rho)$ is the pressure
and the exponent $\gamma$ is called the {\em polytropic index}. The value of
$\gamma$ is determined by the underlying microscopic model. For example,
$\gamma = 3$ stands for non-interacting fermions and the corresponding
analytic solution of the hydrodynamic equations was found by Abanov
\cite{abanov2005hydrodynamics}. Weakly interacting bosons, described by
$\gamma=2$, were recently numerically studied in
Ref.~\cite{yeh2020emptiness}. For the quasi-1D fermions, i.e. 3D fermions confined to 1D by a transverse
harmonic trap, one finds $\gamma = 7/5$ \cite{joseph2011observation}.
Moreover, EFP in Calogero-Sutherland
integrable model
\cite{calogero1969ground,sutherland1971exact,sutherland1972exact} was studied
\cite{franchini2010emptiness}. The leading term in its equation of state has
the polytropic index $\gamma = 3$
\cite{andric1983large,polychronakos1995waves,abanov2005hydrodynamics,stone2008classical},
conforming with the corresponding hydrodynamic solution \footnote{From the hydrodynamic perspective, the difference between the free fermions and the Calogero-Sutherland model is in renormalization of the sound velocity by the interaction parameter, $\lambda$, \cite{kulkarni2009nonlinear}.}.

The goal of this work is to go beyond the above listed examples and find EFP in a 1D polytropic liquid with an arbitrary index. The classical hydrodynamics of polytropic liquids has been attracting attention of mathematicians since 1980's \cite{olver1988hamiltonian,brunelli1997lax,brunelli2004integrable}, when certain instances of classical integrability were 
discovered. Some techniques has been developed for initial condition problems \cite{whitham2011linear,sommerfeld1949partial}  However, the analytical closed form solution was achieved only for some values
of $\gamma$ \cite{kamchatnov2000nonlinear}. Here we utilize these techniques
to construct instanton solutions of EFP for an infinite sequence of rational
indexes. This solution appears in a closed algebraic form, admitting a unique analytic continuation to an arbitrary value of $\gamma\geq 1$.

The sought instanton solution of the hydrodynamic equations of motion involves
distortion of the density in the spatial region of the size of the emptiness,
$\sim R$.  This distortion persists for the time of the order $R/v_s$, where
$v_s$ is the hydrodynamic sound velocity in the liquid.  Therefore the
instanton action, equal to the negative logarithm of EFP, is expected to be
proportional to $R^2/v_s$. These considerations motivate the scaling form of
the leading EFP exponent:
\begin{align}
\lim_{R\to \infty}\frac{-\ln {\cal P}_\mathrm{EFP}(R)}{R^2} = \frac{\rho_0}{\xi}\, f(\gamma),
\label{Eq: leading exponent of EFP}
\end{align}
where the equilibrium density of the 1D liquid $\rho_0$  and  the quantum correlation length $\xi = \hbar/(m
v_s)$ provide  the correct dimensional prefactor  in front of dimesnionless function $f(\gamma)$. 
The speed of sound in this expression is determined by the equation of state in the usual way, 
\begin{align}
 m v_s^2 =  \partial_{\rho} P(\rho)\Big|_{\rho = \rho_0}. 
    \label{Eq: Sound Velocity}
\end{align}
where $P(\rho)$ is the hydrodynamic  pressure and $m$ is mass of the particles. The polytropic equation of state with an exponent $\gamma$ may thus be parametrized as 
 \begin{align}
 P(\rho) = \frac{m v_s^2 }{\gamma \rho_0^{\gamma-1}}\, \rho^\gamma.   
     \label{Eq: Equation of State-pressure}
\end{align} 


\begin{figure}[t] 
    \centering
    \includegraphics[width=0.45\textwidth]{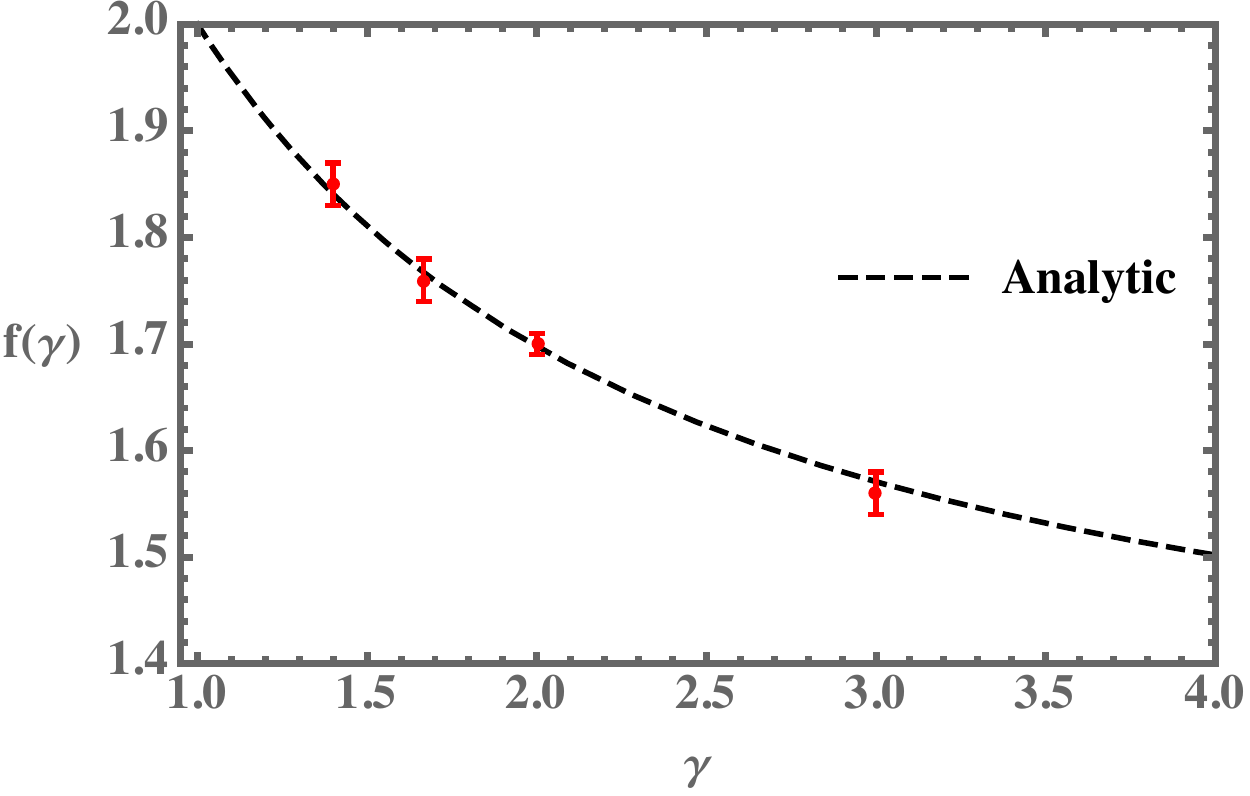}
    \caption{Universal function $f(\gamma)$. The black dashed line is Eq.~(\ref{Eq: Universal Function}), the red symbols are numerical results for $\gamma = 7/5, 5/3, 2, 3$. The numerical results for $\gamma = 2$ is taken from Ref.~\cite{yeh2020emptiness}. }
    \label{fig: Universal Function}
\end{figure}

The analytic expression for the universal function $f(\gamma)$ in Eq.~(\ref{Eq: leading exponent of EFP}) is the main result of this paper. We found:   
\begin{align}
    f(\gamma) =\frac{\pi\ 2^\frac{\gamma - 5}{\gamma - 1}\  \left[\Gamma\left(\frac{\gamma + 1}{\gamma - 1}\right)\right]^2}{ \Gamma\left(\frac{3\gamma - 1}{2\gamma - 2}\right) \left[ \Gamma\left(\frac{\gamma + 1}{2\gamma - 2}\right)\right]^3}\,. \label{Eq: Universal Function} 
\end{align}
where $\gamma \geq 1$. Figure~\ref{fig: Universal Function} shows function $f(\gamma)$ along with numerical results for several values of $\gamma$. In particular, the free fermion point is given by $f(3) = \pi/2$ (and $mv_s = \hbar\pi \rho_0=p_F$ - the Fermi momentum), which agrees with the RMT  \cite{dyson1962statisticalII,mehta2004random} and hydrodynamic \cite{abanov2005hydrodynamics} results. For the weakly interacting bosons  $f(2)=16/(3\pi) \approx 1.698$, which agrees well with the numerical estimate $f(2) = 1.70(1)$ of Ref.~\cite{yeh2020emptiness}. Away from these points 
$f(\gamma)$ is a monotonically decreasing function with the asymptotic limits $f(1) = 2$ and $f(\infty) = 4/\pi$.   

\begin{figure}[t]
    \centering
    \includegraphics[width=0.45\textwidth]{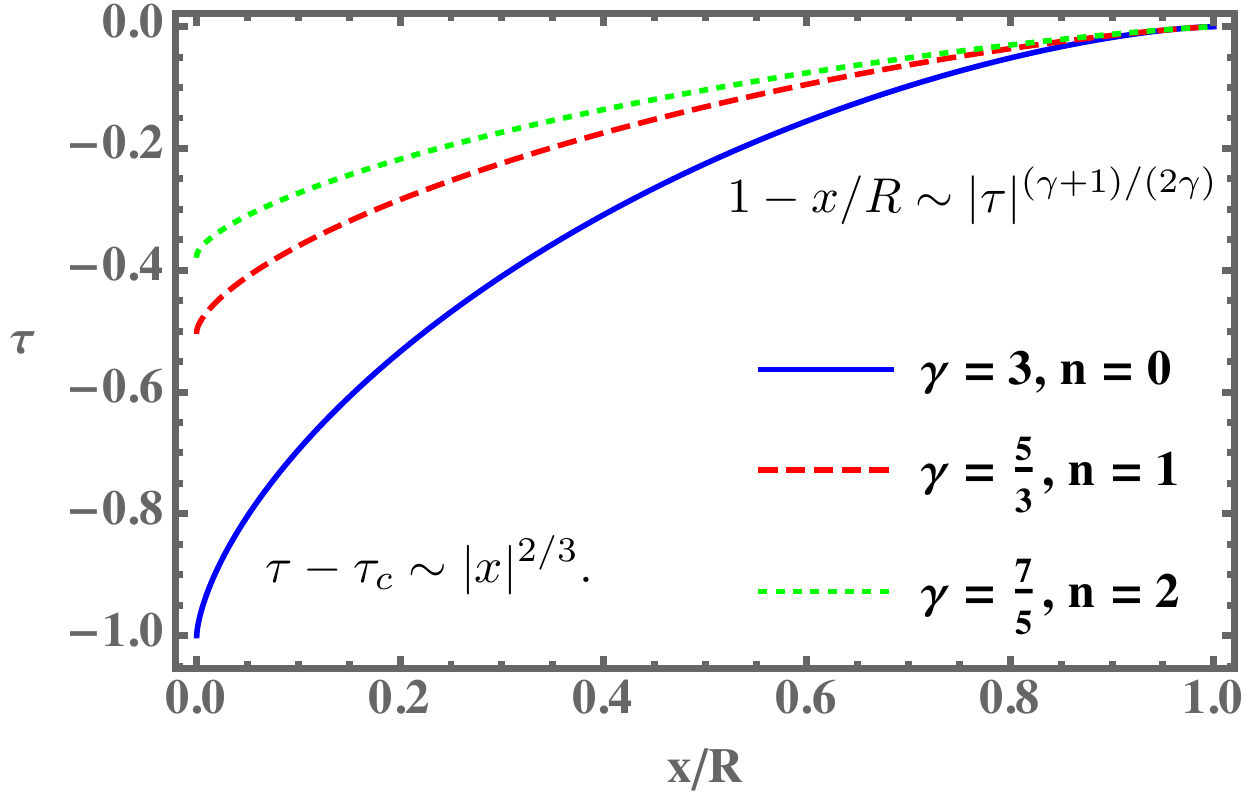}
    \caption{The edge of emptiness region for $\gamma = 3$ (blue solid line), $\gamma =5/3$ (red dashed line) and $\gamma =7/5$ (green dotted line), as given by Eqs.~(\ref{Eq: Emptiness Boundary n=0}), (\ref{Eq: Emptiness Boundary n = 1}) and (\ref{Eq: Emptiness Boundary n = 2}). The asymptotic behaviors are given by Eq. (\ref{Eq: power law around tau = 0 in gamma notation}) and (\ref{Eq: power law around x = 0 in gamma notation}).}
    \label{fig: Emptiness Boundary}
\end{figure}

The appropriate instanton solution exhibits the  empty interval
which nucleates  near $x=0$ at some instance of the imaginary time 
$\tau=-\tau_c\propto R/v_s$. It then develops into
the macroscopic interval $|x|\leq R$ at $\tau=0$ and closes up again at $\tau=+\tau_c$ at
$x=0$. In the free fermion case, $\gamma=3$,
  the emptiness region in $(x,\tau)$ plane is bounded by an astroid
$(x/R)^{2/3}+ (\tau/\tau_c)^{2/3} =1$ \cite{abanov2005hydrodynamics}. For
other $\gamma$'s the exact analytical shape of the empty region is more
complicated. We found that for $\tau \lesssim |\tau_c|$ and $|x|\ll R$, the
emptiness nucleates in the same way as for the free fermions  
\begin{align}
    \tau - \tau_c \sim |x|^{2/3}.
    \label{Eq: power law around x = 0 in gamma notation}
\end{align}
for any $\gamma\geq 1$. However, the other corner of the emptiness region at $|x|\lesssim R$ and $|\tau|\ll \tau_c$ is not 
universal and is described by a $\gamma$-dependent exponent:  
\begin{align}
    R - x \sim |\tau|^{(\gamma+1)/(2\gamma)}. 
    \label{Eq: power law around tau = 0 in gamma notation}
\end{align}
The boundaries of the emptiness region on the $(x,\tau)$ plane for some specific values of $\gamma$ are depicted in Fig.~\ref{fig: Emptiness Boundary}.

The remainder of the paper is organized as follows: in Section \ref{II} we
formulate the instanton approach for calculation of EFP for polytropic liquids
and construct a systematic method to solve hydrodynamic equations. Analytic solutions for a sequence of rational $\gamma$'s are constructed in Section \ref{III}. Conclusions and discussions are presented in Section \ref{IV}. Some technical details are delegated to Appendices.

\section{Instanton solution for  Polytropic Liquids}
\label{II}

\subsection{Instanton Calculus}

The hydrodynamic instanton approach to the emptiness formation, developed in Refs.~\cite{abanov2002probability,abanov2003emptiness,abanov2005hydrodynamics,franchini2005asymptotics,franchini2010emptiness}, is justified in the regime of a macroscopic emptiness, $R\gg \rho_0^{-1} $. A state of the system is characterized by hydrodynamic degrees of freedom: the local particle density, $\rho(x,t)$, and the local current, $j(x,t)$. The two are constrained by the continuity equation, 
\begin{align}
\partial_t \rho + \partial_x j = 0.
\label{Continuity Equation}
\end{align}
The real time  action, that yields proper hydrodynamic equations as its extremal conditions, is given by 
\begin{align}
\label{Eq; Hydro action}
S[\rho,j] = \int\!\!\! \int\! dxdt\, \left[ \frac{mj^2}{2\rho} -  V(\rho)\right]. 
\end{align}
It consists of the liquid's kinetic, $mj^2/(2\rho)$, and internal, $V(\rho)$,
energy densities.  The internal energy density
is related to the pressure through the thermodynamic relation $P(\rho) = \rho\partial_\rho V(\rho) - V(\rho)$. For the polytropic liquid with the pressure, given by Eq.~(\ref{Eq: Equation of State-pressure}) with $\gamma >1$, the internal energy density is thus
\begin{align}
V(\rho) = \frac{m v_s^2 \rho_0}{\gamma-1}\left[ \frac{1}{\gamma} \left(\frac{\rho}{\rho_0}\right)^\gamma   -
\frac{\rho}{\rho_0}\right]. 
\label{Eq: Equation of state}
\end{align}
The linear in $\rho$ term is $-\mu\rho$ with  the chemical potential
$\mu = m v_s^2/(\gamma - 1)$. It fixes the average density to be $\rho_0$
through the condition $\partial_\rho V(\rho)|_{\rho=\rho_0}=0$  and does not affect the equation of state. 
For $\gamma=1$ one finds 
\begin{align}
V(\rho) = m v_s^2 \rho \left[\log(\rho/\rho_0)-1\right]. 
\label{Eq: Equation of state-linear}
\end{align}
Since we are only interested in the leading term
($-\ln {\cal P}_\mathrm{EFP} \sim R^2$) in EFP, we neglect higher gradient terms such as quantum pressure in the equation of state (\ref{Eq: Equation of state}). Including gradient terms
results in sub-leading contributions in $\xi/R \ll 1$.

Variation of the action (\ref{Eq; Hydro action})  over  $\rho$ and $j$ with the continuity constraint, Eq.~(\ref{Continuity Equation}), yields classical Euler equation of the hydrodynamic flow \cite{landau1987fluid}.
The emptiness formation does {\em not} come from the dynamics of this
equation, since the emptiness is a large quantum fluctuation (similar to tunneling), which is outside of the
classically allowed region of the phase space.
In the instanton approach, the path integral $\int {\cal D}\rho {\cal D}j
e^{iS[\rho,j]/\hbar}\delta(\partial_t \rho + \partial_x j )$, with proper boundary
conditions, determines the quantum transition amplitude. One has to
  deform the fields  into the complex plane to go through a
  classically forbidden stationary configuration corresponding to the emptiness. The probability of such rare event is  ${\cal P}\propto |e^{iS_{\mathrm{inst}}/\hbar}|^2$, where the classical action along the instanton trajectory, $S_{\mathrm{inst}}$, has a positive imaginary part.

  After the Wick rotation, $t \rightarrow -i\tau$, to the imaginary time, the
  equations of motions are solved by a real $\rho$ and a purely imaginary $j$
  (i.e. the integration contour over $j$ should be deformed to pass through
  the imaginary saddle point). It is convenient thus to redefine
  $j \rightarrow ij$ such that the saddle point solutions for both $\rho$ and
  $j$ are real functions. Finally, we pass to dimensionless coordinates and
  fields:
  $x \rightarrow Rx,\ \tau \rightarrow (R/v_s)\tau,\ \rho \rightarrow \rho_0
  \rho$ and $j \rightarrow (\rho_0 v_s)j$. The corresponding Eucledian action
  is
\begin{align}
\label{Eq: Euclidean action rho j}
&\frac{i}{\hbar}\,S[\rho,j] = -\frac{\rho_0 R^2}{\xi}\int\!\!\! \int\! dxd\tau\, \Big[ \frac{j^2}{2\rho} + V(\rho)\Big];\\
&V(\rho) = \frac{1}{\gamma - 1}\left[\frac{\rho^\gamma}{\gamma} - \rho\right] .
\label{Eq: Euclidean action V}
\end{align}
The Planck constant is suppressed below. 
The internal energy density $V(\rho)$  changes  sign
(cf. Eq.~(\ref{Eq; Hydro action})) similarly
to the inverted potential in the tunneling problem. The dimensionless equations of motion in the imaginary time are
\begin{align}
&\partial_\tau \rho + \partial_x (\rho v) = 0 ; \label{Eq: Hydrodynamic Eq-1}\\
&\partial_\tau v + v \partial_x v = \rho^{\gamma - 2} \partial_x \rho,
\label{Eq: Hydrodynamic Eq-2}
\end{align}
where the velocity field, $v(x,\tau)$,  is defined as $v = j/\rho$. 
The fact that the dimensionless equations of motion and the boundary conditions (see below) depend only on $\gamma$ and no other parameters, justifies the scaling form (\ref{Eq: leading exponent of EFP}) of EFP.

Apart from the emptiness formation, the polytropic hydrodynamic equations (\ref{Eq: Hydrodynamic Eq-1}) and (\ref{Eq: Hydrodynamic Eq-2}) appear in a variety of distinct fields across physics. From early studies of Cauchy (initial condition) problem of unstable media \cite{trubnikov1987unstable} and large-N limit of matrix model \cite{matytsin1994large} to the recent works on large deviation in classical stochastic systems \cite{vilenkin2014extreme,meerson2014extreme,Alex2016short-time,smith2018landau}. 

\subsection{Riemann Invariants and Hodograph Transformation}

Following  Ref.~\cite{kamchatnov2000nonlinear},  Eqs. (\ref{Eq: Hydrodynamic Eq-1}) and (\ref{Eq: Hydrodynamic Eq-2}) may be reformulated to simplify their solution. First, one introduces the Riemann invariant, $\lambda(x,t)$, and the complex ``velocity",  $w(x,t)$, as
\begin{align}
&\lambda = v+i \tfrac{2}{\gamma-1} \rho^\frac{\gamma-1}{2}, \label{Eq: Definition of lambda}\\
&w = v + i \rho^\frac{\gamma-1}{2}.
\end{align}
In terms of these quantities the equations of motion (\ref{Eq: Hydrodynamic
  Eq-1}), (\ref{Eq: Hydrodynamic Eq-2}) acquire a more symmetric form: 
\begin{align}
&\partial_{\tau} \lambda + w \partial_x \lambda =0,\label{Eq: Riemann Invariants-1}\\
&\partial_{\tau} \overline{\lambda} + \overline{w}\partial_x \overline{\lambda}=0,\label{Eq: Riemann Invariants-2}
\end{align}
where $\overline{\lambda}$   and $\overline{w}$  are complex conjugates of $\lambda$ and  $w$.  Notice that for 
$\gamma=3$, $\lambda = w=v+i\rho$, which reduces the problem to
  finding  an analytic function $\lambda(x,\tau)$ which can be done
in a relatively straightforward way \cite{abanov2005hydrodynamics}.  

To proceed in the general case we employ the so-called Hodograph transformation (See Appendix \ref{Appendix: Hodograph}). The idea is to find $x(\lambda, \overline{\lambda})$ and $\tau(\lambda, \overline{\lambda})$ as functions of Riemann invariants.  Under the Hodograph transformation Eqs.~(\ref{Eq: Riemann Invariants-1}) and (\ref{Eq: Riemann Invariants-2}) become
\begin{align}
&\partial_{\overline{\lambda}}x-w\partial_{\overline{\lambda}}\tau = 0,\label{Eq: Hodograph-1}\\
&\partial_{\lambda}x-\overline{w}\partial_{\lambda}\tau = 0.\label{Eq: Hodograph-2}
\end{align}
Now, one can solve these equations of motion by adopting the ansatz
\begin{align}
x-w\tau = \partial_{\lambda} \mathcal{V},\label{Eq: Ansatz-1}\\
x-\bar{w}\tau = \partial_{\overline{\lambda}} \mathcal{V},\label{Eq: Ansatz-2}
\end{align}
where the real function $\mathcal{V}(\lambda,\bar{\lambda})$ depends only on $\lambda$ and $\overline{\lambda}$. After substituting this ansatz into the equations of motion (\ref{Eq: Hodograph-1}) and (\ref{Eq: Hodograph-2}), one obtains the equation for the function $\mathcal{V}(\lambda,\bar{\lambda})$,
\begin{align}
\partial_{\lambda}\partial_{\overline{\lambda}} \mathcal{V} = \frac{n}{\lambda-\overline{\lambda}}\, \left(\partial_{\lambda} \mathcal{V} - \partial_{\overline{\lambda}} \mathcal{V}\right), \label{Eq: electrostatic-like V}
\end{align}
where $n$ is defined as 
\begin{align}
\label{Eq: gamma n relation}
n=-\frac{1}{2} \frac{\gamma -3}{\gamma-1};\qquad\qquad  \gamma = \frac{2n+3}{2n+1}.
\end{align}
The Riemann invariant in terms of $n$ is
\begin{align}
    \lambda = v + i(2n+1)\rho^{1/(2n+1)}.
    \label{Eq: lambda in terms of n}
\end{align}
The main idea of this mathematical manipulation is to map the original
hydrodynamic equations, (\ref{Eq: Hydrodynamic Eq-1}) and (\ref{Eq:
  Hydrodynamic Eq-2}), onto an electrostatic-like problem of finding a
solution for the 2D ``potential" $\mathcal{V}$ in the complex $\lambda$
plane.  The emptiness condition only partially fixes the density at $\tau = 0$,
i.e. $\rho (|x| < 1,\tau =0) = 0$, but the velocity is left unspecified  at $\tau
= 0$. This seems to provide insufficient information to find the
potential. However, in terms of Riemann invariants, both density and velocity
are combined together. The boundary conditions for the  density is actually
also constraining the velocity as well by  requiring
analyticity of the potential in the plane of Riemann invariants.

The method of Riemann invariants has been known for a while, (see Ref.~\cite{whitham2011linear} and references there) and it is a powerful tool. It was deployed in e.g. recent studies of Bose liquid \cite{Isoard2019wave} and relativistic fluid \cite{kamchatnov2019landau}. It was noticed by Kamchatnov \cite{kamchatnov2000nonlinear} that
Eq.~(\ref{Eq: electrostatic-like V}) admits a closed form  analytic solution if $n$ is a
{\em non-negative integer}.  Employing this approach along with the relation,
found by Abanov \cite{abanov2005hydrodynamics}, between EFP and the asymptotic
behavior of the density at $x \rightarrow \infty$ and $\tau = 0$, the EFP may
be calculated exactly for the discrete sequence of the rational polytropic
indices, Eq.~(\ref{Eq: gamma n relation}). Finally, this result allows for the
unique analytic continuation to find EFP for any $\gamma\geq 1$.

\subsection{Instanton solution for integer $n$}
Let's first examine the simplest case, $n = 0$, where the right hand side of
Eq. (\ref{Eq: electrostatic-like V}) vanishes. Therefore $\mathcal{V}$ is
given by the sum of two arbitrary analytic functions
\begin{align}
\mathcal{V} = F_0(\lambda)+G_0(\overline{\lambda}),\qquad \text{for $n = 0$.}
\end{align}
For $n \neq 0$, the right hand side of Eq. (\ref{Eq: electrostatic-like V})
complicates the solution by introducing  coupling between $\lambda$ and $\overline{\lambda}$. The structure of Eq. (\ref{Eq: electrostatic-like V}) with integer $n$ suggests to look for its solution in the form of  a  series expansion \cite{kamchatnov2000nonlinear}
\begin{align}
\mathcal{V} = \frac{F_0(\lambda)+G_0(\overline{\lambda})}{(\lambda-\overline{\lambda})^n} + \sum_{m=1}^\infty \mathcal{V}_m,
\end{align}
where $\{ \mathcal{V}_m \}$ are to  be determined order by order. By substituting it into Eq. (\ref{Eq: electrostatic-like V}), one finds
\begin{align}
    &n(n-1)\frac{F_0(\lambda)+G_0(\overline{\lambda})}{(\lambda -\overline{\lambda})^{n+2}}+ \sum_{m=1}^\infty \Big[ \partial_\lambda \partial_{\overline{\lambda}} \mathcal{V}_m - \frac{n}{\lambda -\overline{\lambda}}(\partial_\lambda \mathcal{V}_m -\partial_{\overline{\lambda}} \mathcal{V}_m)  \Big] = 0.
\end{align}
In order to cancel the term with $(\lambda -\overline{\lambda})^{-(n+2)}$, one requires $\mathcal{V}_1$ to have the form
\begin{align}
    \mathcal{V}_1 = a_1 \frac{F_1(\lambda) -G_1(\overline{\lambda})}{(\lambda -\overline{\lambda})^{n+1}},
\end{align}
where $a_1$ is an overall coefficient and $F_1$ and $G_1$ are analytic functions. The requirement of cancellation of $(\lambda -\overline{\lambda})^{-(n+2)}$ term leads to the recurrence relation: $a_1 = -n(n-1)$, $F_0 = \partial_\lambda F_1$ and $G_0 = \partial_{\overline{\lambda}}G_1$. Now,  Eq. (\ref{Eq: electrostatic-like V}) becomes
\begin{align}
    &a_1(n+1)(n-2)\frac{F_1(\lambda)-G_1(\overline{\lambda})}{(\lambda - \overline{\lambda})^{n+3}}+ \sum_{m=2}^\infty \Big[ \partial_\lambda \partial_{\overline{\lambda}} \mathcal{V}_m - \frac{n}{\lambda -\overline{\lambda}}(\partial_\lambda \mathcal{V}_m -\partial_{\overline{\lambda}} \mathcal{V}_m)  \Big] = 0,
\end{align}
where the term with $(\lambda -\overline{\lambda})^{-(n+3)}$ is left to be cancelled by a proper choice of $\mathcal{V}_2$.  By repeating the corresponding cancellation  procedure, subsequent $\{ \mathcal{V}_m \}$ are recovered order by order. 

The series expansion terminates if $n$ is an integer. For example, for $n = 1$, $a_1=0$ and thus
\begin{align}
\mathcal{V} = \frac{F_0(\lambda)+G_0(\overline{\lambda})}{\lambda-\overline{\lambda}}, \qquad \text{for $n = 1$,}
\end{align}
which can be verified by a direct substitution in Eq. (\ref{Eq:
  electrostatic-like V}). For a positive integer $n$, there are exactly $n$
terms of series expansion in terms of $(\lambda - \overline{\lambda})$, as
follow
\begin{align}
\mathcal{V} = \frac{F_0(\lambda)+G_0(\overline{\lambda})}{(\lambda-\overline{\lambda})^n} + \sum_{m=1}^{n-1} a_m\frac{F_m(\lambda)+(-1)^m G_m(\overline{\lambda})}{(\lambda-\overline{\lambda})^{n+m}},\label{Eq: V ansatz}
\end{align}
where all the $\{ F_m \}$ functions only depend on $\lambda$ and all the $\{ G_m \}$ functions only depend on $\overline{\lambda}$. The $\{ a_m \}$ are the coefficients of series expansion. The recurrence relations for $F_m$ and $G_m$  functions are 
\begin{align}
&F_{m-1} = \partial_{\lambda} F_m, \label{Eq: recurrence F}\\
&G_{m-1} = \partial_{\overline{\lambda}} G_m, \label{Eq: recurrence G}
\end{align}
and for the coefficients
\begin{align}
a_1 &= -n(n-1), \label{Eq: recurrencea1}\\
a_m &= -\frac{1}{m}a_{m-1}(n+m-1)(n-m).\label{Eq: recurrenceam}
\end{align}
The series terminates since $a_m = 0$ for $m \geq n$.

The specific form of $\{ F_m \}$ and $\{ G_m \}$ functions is to be determined
from the boundary conditions.  To find the boundary conditions for
$\mathcal{V}$ one needs to go back to the original hydrodynamic variables and
discuss the boundary conditions for the density and the velocity. For
simplicity, let us focus on $x > 0$ in the following. Solutions at $x < 0$ can
be obtained by spatial inversion: $\rho (x,\tau) = \rho (-x,\tau)$ and
$v(x,\tau) = -v(x,\tau)$. The instanton solution evolves from a uniform state
at a distant past, $\tau=-\infty$, to a state with the emptiness, i.e. zero
density for $|x| < 1$, at the observation time, $\tau=0$. At $\tau = 0$, the
density diverges at $x = 1$ since the displaced particles accumulate on the emptiness
boundary. In terms of Riemann invariants, $|\lambda| \rightarrow \infty$ at
$x = 1$ and $\tau = 0$ and from (\ref{Eq:
  Ansatz-1}) we have 
\begin{align}
    \partial_\lambda \mathcal{V} \Big|_{|\lambda| \rightarrow \infty} = 1,\label{Eq: Density diverges at x = R}
\end{align}
which fixes the boundary of emptiness $x = 1$. Far away from the emptiness,
the density decays to the average density and the velocity approaches zero. In
general, the density decays like $(\rho - 1) \propto 1/x^2$ as
$x \rightarrow \infty$ \cite{abanov2005hydrodynamics}. In terms of Riemann
invariants, $\lambda \rightarrow i\frac{2}{\gamma-1} = i(2n+1)$ as
$x \rightarrow \infty$.  Substituting this condition into Eq. (\ref{Eq:
  Ansatz-1}), one finds the other boundary condition:
\begin{align}
\partial_\lambda \mathcal{V} \Big|_{\lambda \rightarrow i(2n+1)} \sim \frac{1}{\sqrt{\lambda^2+(2n+1)^2}}\label{Eq: Sqrt Root divergence},
\end{align}
where the square root divergence on the right hand side comes from the $1/x^2$
behavior of the density. The requirement of zero density is encoded in the boundary
conditions (\ref{Eq: Density diverges at x = R}) and (\ref{Eq: Sqrt Root
  divergence}). Indeed, they imply that the solution has a branch point at $x
= 1$ for $\tau = 0$ so that $\lambda$ becomes  
a purely real function at $x < 1,\tau = 0$ corresponding to $\rho=0$.

With the boundary condition (\ref{Eq: Density diverges at x = R}) and (\ref{Eq: Sqrt Root divergence}) and the recurrence relations (\ref{Eq: recurrence F}) and (\ref{Eq: recurrence G}) one can construct all the $\{ F_m \}$ and $\{ G_m \}$ functions as long as the last terms  $F_{n-1}$ and $G_{n-1}$ are specified. For the special case, $n = 0$,
\begin{align}
&F_0 =  \sqrt{\lambda^2+1},\qquad\qquad 
&G_0 = \overline{F}_0,
\end{align}
where  the fact that $\mathcal{V}$ is a real function is employed. For a
positive integer $n$, the series of $\mathcal{V}$
terminates at the term with $F_{n-1}$, which must be taken as 
\begin{align}
&F_{n-1} = \frac{\lambda}{n!} \Big[ \lambda^2+(2n+1)^2 \Big]^{\frac{2n-1}{2}}, \label{Eq: Boundary condition of last term F}\\
&G_{n-1} = (-1)^n \overline{F}_{n-1},
\label{Eq: Boundary condition of last term G}
\end{align}
where the coefficient $1/n!$ results from  Eq. (\ref{Eq: Density diverges at x
  = R}). Finally, one may find all  $\{ F_m \}$ functions, $\{ G_m \}$
functions and coefficient $\{ a_m \}$ from the recurrence relations, Eqs.~(\ref{Eq: recurrence F}),(\ref{Eq: recurrence G}),(\ref{Eq: recurrencea1}),(\ref{Eq: recurrenceam}).

\subsection{Emptiness formation probability} 

We are now at the position to calculate EFP. The semiclassical transition amplitude is given by $e^{iS_{\mathrm{inst}}(R)}$. Since EFP is a probability of the fluctuation with respect to the ground state, we need to normalize this amplitude by dividing it by $e^{iS_0}$, where $S_0$ is the action evaluated at the static ground state solution, $\rho = \rho_0$ and $j = 0$. This results in the EFP of the form  
\begin{align}
\label{EFP Sopt}
-\ln {\cal P}_\mathrm{EFP}(R) = 2\text{Im} \left[S_{\mathrm{inst}}(R)-S_0\right] = \frac{\rho_0 R^2}{\xi} \, f(n),
\end{align}
where the second equality used the rescaled action Eq. (\ref{Eq: Euclidean action rho j}) and $f(n)$ is a function depending on $n$ only. 

Following Ref.~\cite{abanov2005hydrodynamics}, one may connect EFP with the
asymptotic behavior of the density at large $x$ at $\tau=0$.   We rederive
this relation  in  Appendix \ref{Appendix: Analytic formula for EFP} and here only quote the result
\begin{align}
   i \partial_{\rho_0} \left[S_{\mathrm{inst}}-S_0\right] = \frac{\pi R^2 \alpha}{\xi} ,
    \label{Eq: derivative lnP and alpha}
\end{align}
where $\alpha$ is a coefficient in the following generic asymptotic expansion of the density, $\rho(x \rightarrow \infty,\tau =0)$,
\begin{align}
    \rho(x,0) = 1 + \frac{\alpha}{x^2} + \mathcal{O}\left(\frac{1}{x^4}\right).
    \label{Eq: asymptotics of density at x goes to infty}
\end{align}
For polytropic liquids, the correlation length depends on the average density as $1/\xi \propto \rho_0^{(\gamma-1)/2}$ $=\rho_0^{1/(2n+1)}$. After integrating over $\rho_0$, one finds 
\begin{align}
 i\left[  S_{\mathrm{inst}}(R)-S_0\right] = \frac{\pi \rho_0 R^2}{\xi} \,\frac{2n+1}{2n+2}\, \alpha.
   \label{Eq: na fig}
\end{align}
To determine the $n$ dependence of the coefficient $\alpha$, we employ the instanton solution for positive integer $n$. According to Eq. (\ref{Eq: Ansatz-1}), (\ref{Eq: Boundary condition of last term F}) and (\ref{Eq: Boundary condition of last term G}), the solution for any $n$ can be explicitly written down. The leading term  at $x \rightarrow \infty$  comes solely from $\partial_\lambda F_0(\lambda)=\partial_\lambda^n F_{n-1}(\lambda)$ due to the boundary condition (\ref{Eq: Sqrt Root divergence}) and the  recurrence relation (\ref{Eq: recurrence F}). This way one arrives at
\begin{align}
    x-w\tau = \frac{1}{(\lambda-\overline{\lambda})^n} \partial_\lambda^n \left\{  \frac{\lambda}{n!} \Big[ \lambda^2+(2n+1)^2 \Big]^{\frac{2n-1}{2}} \right\}+...,
\end{align}
where the sub-leading terms are omitted. To satisfy the boundary condition (\ref{Eq: Sqrt Root divergence}) and thus generate the $\left[ \lambda^2 +(2n+1)^2 \right]^{-1/2}$ term, all $n$ derivatives should act on the square bracket term in this expression. As a result one finds  
\begin{align}
    x-w\tau =
    \frac{(2n-1)!!}{n!}\frac{\lambda^{n+1}}{(\lambda-\overline{\lambda})^n\sqrt{\lambda^2+(2n+1)^2}}+... .
\end{align}
At $\tau = 0$, take the limit: $x \rightarrow \infty$, $\rho \rightarrow 1$, \mbox{$\lambda \rightarrow i(2n+1)\rho^{1/(2n+1)}$} and $\overline{\lambda} \rightarrow -i(2n+1)\rho^{1/(2n+1)}$,  as in Eq.~(\ref{Eq: lambda in terms of n}); and  
\mbox{$\left[ \lambda^2 + (2n+1)^2 \right] = (2n+1)^2$} \mbox{$\times (1-\rho^{2/(2n+1)}) \approx -2(2n+1)(\rho - 1)$}. Therefore 
\begin{align}
    x \approx \frac{(2n+1)!!}{2^n n!} \frac{1}{\sqrt{2(2n+1)(\rho - 1)}}.
\end{align}
Accordingly, the coefficient $\alpha$ in Eq.~(\ref{Eq: asymptotics of density at x goes to infty})   is given by
\begin{align}
    \alpha = \frac{1}{2(2n+1)} \left[ \frac{(2n+1)!!}{2^n n!} \right]^2 ,
\end{align}
and  EFP is found with the help of Eqs.~(\ref{EFP Sopt}) and (\ref{Eq: na fig}) as
\begin{align}
    -\ln {\cal P}_\mathrm{EFP}(R) = \frac{\rho_0 R^2}{\xi} \frac{\pi}{2n+2}\left[ \frac{(2n+1)!!}{2^n n!} \right]^2.
\end{align}
Finally, the function $f(n)$ may be written using the gamma-function representation of the factorials:   
\begin{align}
    f(n) = \frac{\pi \Gamma^2(2n+2)}{2^{4n+1} \Gamma(n+2) \Gamma^3(n+1)}.
    \label{Eq:vse-kozly}
\end{align}
This expression can be extended to real  $n$  due to the uniqueness of the analytic continuation of the gamma function according to Bohr-Mollerup theorem \cite{artin2015gamma}. Converting $n$ into $\gamma$ with the help of Eq.~(\ref{Eq: gamma n relation}), one arrives at Eq. (\ref{Eq: Universal Function}). 

To illustrate the validity of Eq.~(\ref{Eq: derivative lnP and alpha}), we numerically calculate  $f(n)$ by substituting the instanton solutions for $n = 0$, 1 and 2 into Eq. (\ref{Eq: Euclidean action rho j}) and performing Monte
Carlo integration.  The numerical results are summarized in
Table. \ref{table:1}.

\begin{table}[h!]
    \centering
    \begin{tabular}{|c|c|c|c|}
    \hline
     &  \ $n = 0\ (\gamma = 3)$ \ & \ $n = 1\ (\gamma = 5/3)$ \ & \ $n = 2\ (\gamma = 7/5)$ \ \\[1pt]
    \hline
    $f(n)$ & 1.56 $\pm$ 0.02   & 1.76 $\pm$ 0.02   & 1.85 $\pm$ 0.02  \\
    \hline
    Eq.(\ref{Eq:vse-kozly})& $\pi/2\approx 1.571 $ & $9\pi/16\approx 1.767 $ & $75\pi/128\approx 1.841$\\
    \hline
    \end{tabular}
    \caption{Numerical value of function $f(n)$ for $n = 0, 1, 2$. }
    \label{table:1}
\end{table}

\section{Shape of the empty region}
\label{III}

Having found the systematic way to construct analytic solutions for polytropic
fluid, we now explicitly write them down for $n = 0, 1$ and $2$.  The $n = 0$
case  corresponds to the free fermions ($\gamma = 3$) and is the simplest case
of the polytropic fluid since the Riemann invariants $\lambda$ equals to its
complex ``velocity" $w$. In other words, Eq. (\ref{Eq: Riemann Invariants-1})
and (\ref{Eq: Riemann Invariants-2}) become a pair of  complex conjugate
Hopf equations whose closed form solution is a generic analytic function \cite{abanov2005hydrodynamics}. This property can be understood from Eq. (\ref{Eq: electrostatic-like V}), where the right hand side vanishes at $n = 0$ and $\mathcal{V}(\lambda, \overline{\lambda})$ consists of two analytic functions: one depends only on $\lambda$ and the other -- only  on $\overline{\lambda}$. Therefore, 
\begin{align}
&\mathcal{V} = F_0 + G_0,\qquad\qquad 
&F_0 = \sqrt{\lambda^2 + 1},
\end{align}
where $G_0 = \overline{F}_0$, since $\mathcal{V}$ is real. 
For $n = 1$  the corresponding solution is
\begin{align}
&\mathcal{V} = \frac{F_0+G_0}{\lambda - \overline{\lambda}},\qquad\qquad
&F_0 = \lambda\sqrt{\lambda^2+9},
\end{align}
where $G_0 = - \overline{F}_0$. For $n = 2$: 
\begin{align}
&\mathcal{V} = \frac{F_0+G_0}{(\lambda-\overline{\lambda})^2} - 2\frac{F_1-G_1}{(\lambda-\overline{\lambda})^3},\qquad
&F_1=\frac{\lambda}{2}(\lambda^2+25)^{3/2},
\end{align}
where $F_0 = \partial_\lambda F_1$ and $G_{0,1} = \overline{F}_{0,1}$. 

\begin{figure}[t]
    \centering
    \includegraphics[width=0.45\textwidth]{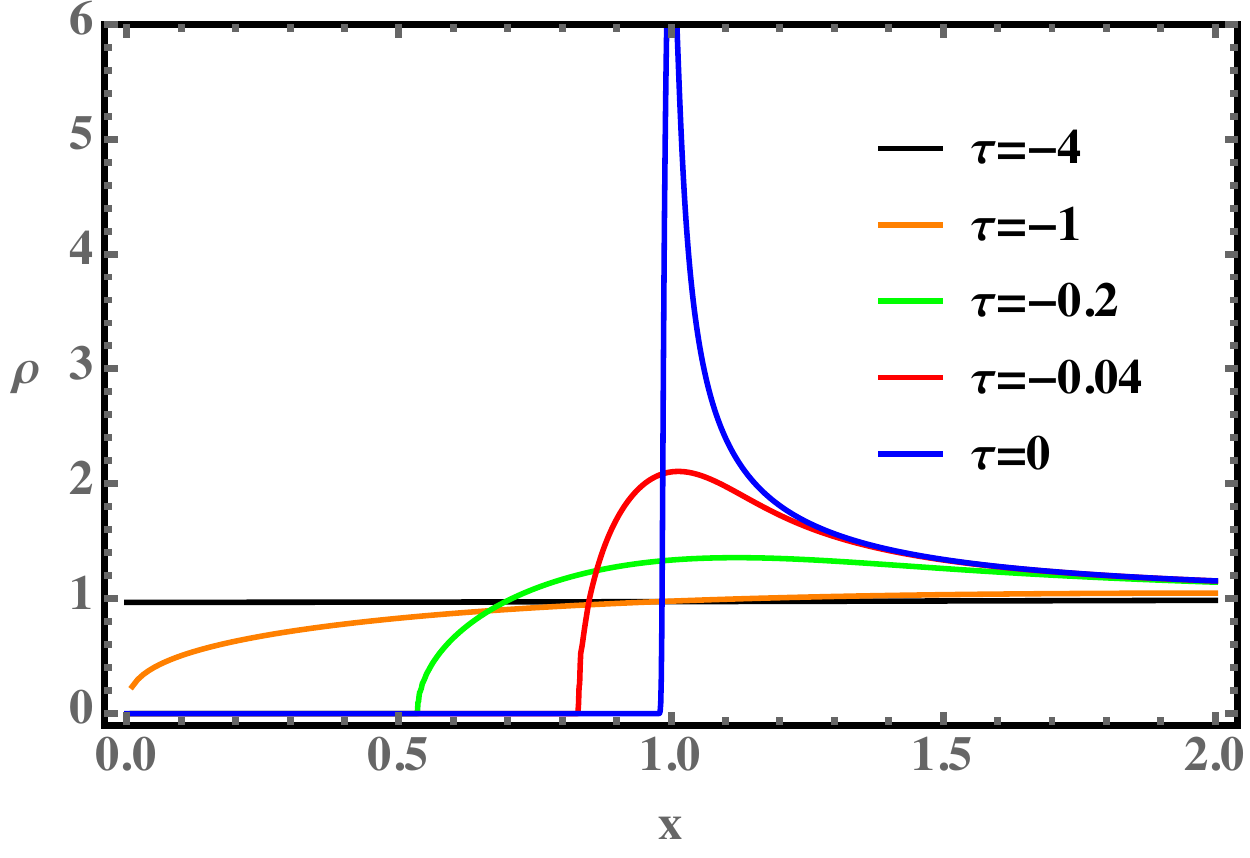}
    \includegraphics[width=0.45\textwidth]{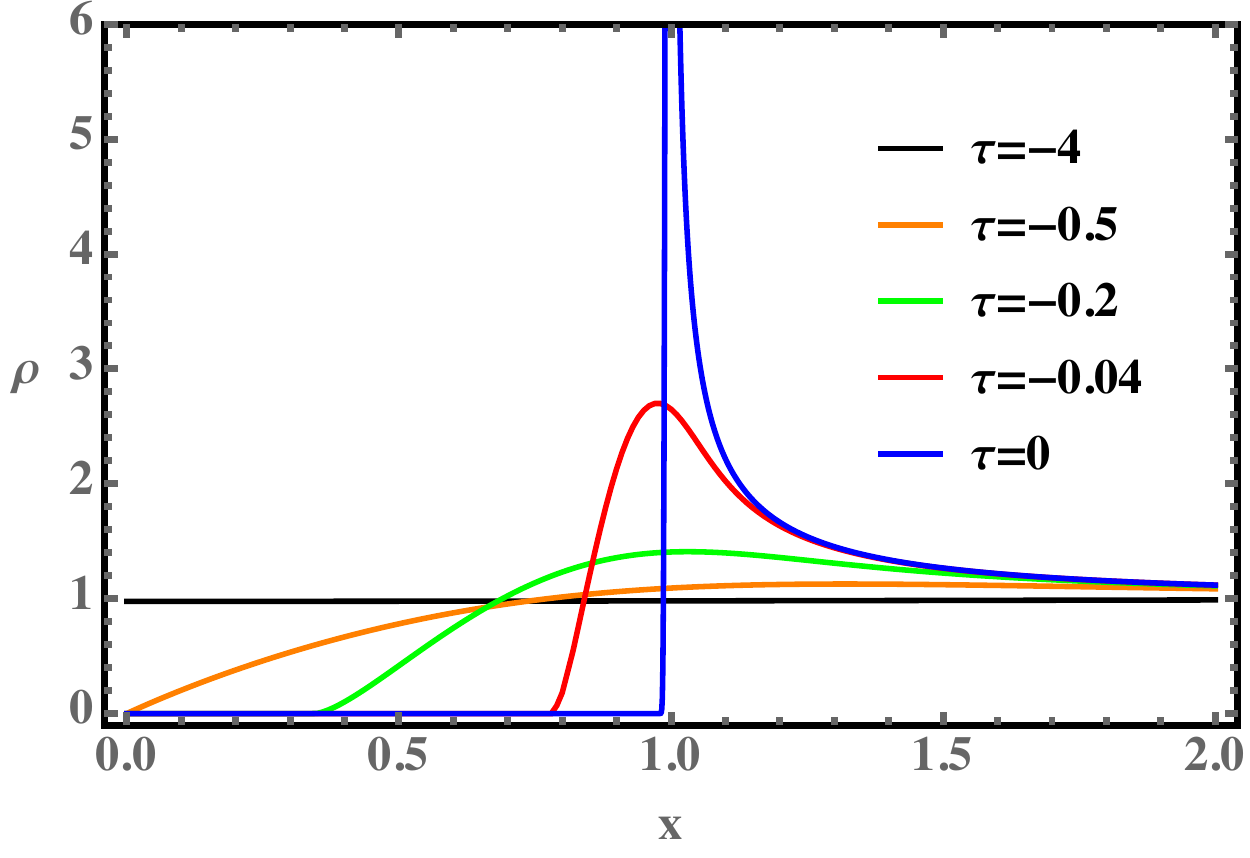}
    \includegraphics[width=0.45\textwidth]{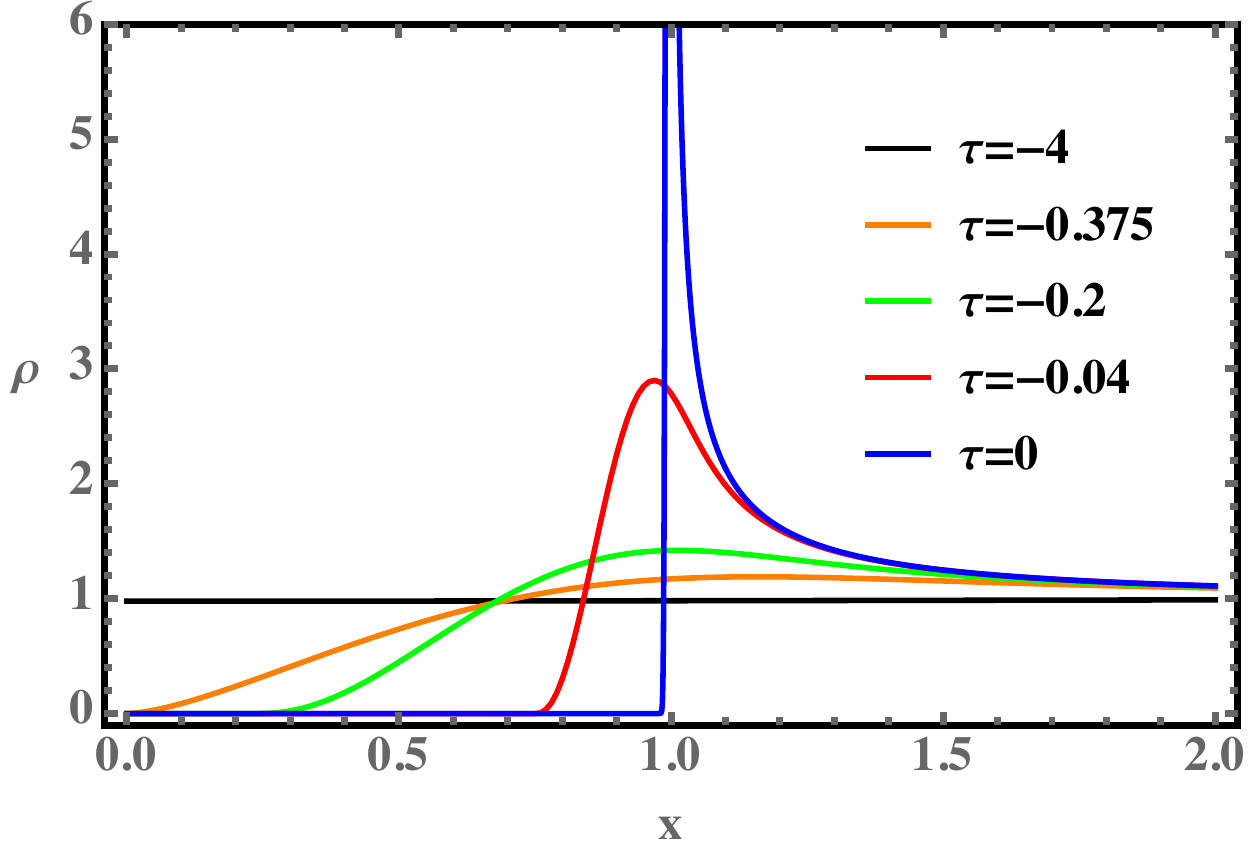}
    \caption{Time evolution of the density $\rho(x,\tau)$ for $n = 0$ (upper left), $n=1$ (upper right) and $n=2$ (bottom). The density evolves from the uniform value, $\rho = 1$, at large negative $\tau$  towards the emptiness within $|x| < 1$ at $\tau = 0$.}
    \label{fig: Density}
\end{figure}

The imaginary time evolution of the density for these three cases is
shown  in Fig. \ref{fig: Density}. It is determined by numerically solving
algebraic Eqs. (\ref{Eq: Ansatz-1}) and (\ref{Eq: Ansatz-2})  with the above
explicit expressions for $\mathcal{V}$. The fluid evolves from the uniform
density into the emptiness profile in Fig. \ref{fig: Density}. There is an
empty region in the $(x,\tau)$ plane, where the density is zero. To find the
boundary of this region we note that for $\rho =0$ the Riemann invariants are
degenerate, $\lambda=\bar\lambda = v$ and our solution represents an equation
for $v(x,t)$. For $n = 0, 1, 2$  we have
\begin{align}
&x - v\tau = \frac{v}{\sqrt{v^2+1}}, \label{Eq: Equation of Emptiness Region n=0}\\
&x - v\tau = \frac{3v}{2\sqrt{v^2+9}}-\frac{v^3}{2(v^2+9)^{3/2}},\label{Eq: Equation of Emptiness Region n=1}\\
&x - v\tau = \frac{15v}{8\sqrt{v^2+25}}-\frac{5v^3}{4(v^2+25)^{3/2}} + \frac{3v^5}{8(v^2+25)^{5/2}}.\label{Eq: Equation of Emptiness Region n=2}
\end{align}
Generally speaking the velocity $v$ is a multi-valued function of coordinate
$x$ parametrised by $\tau$. The ends of the emptiness interval for a given
$\tau$ are real values $x=x_\pm(\tau)$  where two branches of $v$
meet. In other words $x_\pm$ are maximum and minimum value of $x$ as function
of $v$ at given $\tau$ obtained from  
Eqs.~(\ref{Eq: Equation of Emptiness Region n=0}),(\ref{Eq: Equation of
 Emptiness Region n=1}),(\ref{Eq: Equation of Emptiness Region n=2}).
Such real  points exist only within the interval $-\tau_c<\tau<\tau_c$ (e.g.,
$\tau_c = 1$ for $n = 0$). Outside this interval the branch points move away
from the real axis of $x$.

\begin{figure}[h!]
    \centering
    \includegraphics[width=0.35\textwidth]{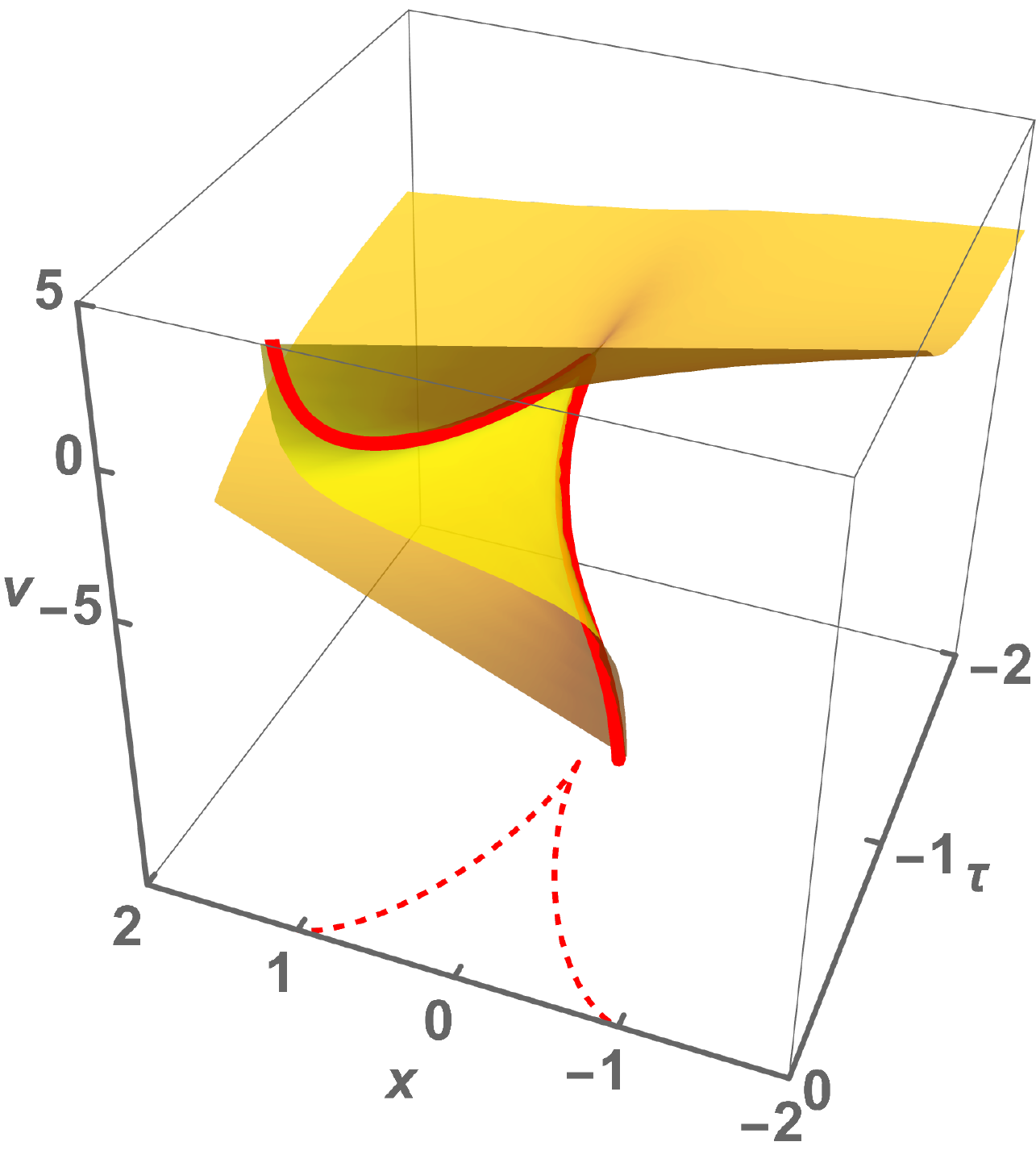}
    \caption{A 2D surface of Eq. (\ref{Eq: Equation of Emptiness Region n=0}). The red (solid) curve on the surface shows the local maximum (minimum) value of $x$ which indicates the boundary of the empty region. After $\tau = -1$, $v$ is a multi-valued function. The red (dashed) curve shows the projection onto $(x,\tau)$ plane is described by Eq. (\ref{Eq: Emptiness Boundary n=0}).}
    \label{fig: 2D surface}
\end{figure}

To visualize this construction
the $n = 0$ case (\ref{Eq: Equation of Emptiness Region n=0}) is plotted in
Fig. \ref{fig: 2D surface}.
The red solid curve on the surface in Fig. \ref{fig: 2D surface} are the collection of these local minima
and maxima. The emptiness boundary is the projection of the red
solid curve onto the space-time plane along the $v$ axis, depicted by the red dashed curve in
Fig. \ref{fig: 2D surface}. For other values of $n$ the emptiness boundary is obtained
by the same procedure and is plotted in Fig. \ref{fig: Emptiness Boundary}.

Let's examine the time around the beginning of the emptiness formation, where
one can do small $v$ expansion for Eq. (\ref{Eq: Equation of Emptiness Region
  n=0}), (\ref{Eq: Equation of Emptiness Region n=1}) and (\ref{Eq: Equation
  of Emptiness Region n=2}). For any $n$  the solutions can be approximated by a cubic polynomial:
\begin{align}
    x \approx (\tau - \tau_c)v - b_n v^3,
    \label{Eq: cubic polynomial equation around x = 0}
\end{align}
where $b_n$ is a positive constant depending on $n$. From Eq.~(\ref{Eq: cubic
  polynomial equation around x = 0}) it follows that $v$ is multi-valued and
$x$ has local maximum and minimum for $\tau >\tau_c$. The power law behavior
of the emptiness boundary is determined from the position of the local minimum
and maximum of $x$ on $(x,\tau)$ plane. According to Eq. (\ref{Eq: cubic
  polynomial equation around x = 0}), one finds
\begin{align}
    \tau - \tau_c \propto |x|^{2/3}.
    \label{Eq: Power law near the beginning of emptiness}
\end{align}
This explains the universal power law exponent 2/3 at the start of the
emptiness as a consequence of an underlying cubic
equation. This scenario of  emptiness formation can be described as a cusp
catastrophe in the catastrophe theory \cite{arnol2003catastrophe}, where the
cusp catastrophe is classified as $A_3$ group and the cusp has the exponent
2/3 \cite{berry1980iv}. Moreover, the Burgers' equation features  the same
exponent 2/3 in the shock wave formation \cite{bessis1984pole}.

Focusing now on the vicinity of $\tau=0$ and point $x = 1$, one can perform large $v$ expansion for Eq. (\ref{Eq: Equation of Emptiness Region n=0}), (\ref{Eq: Equation of Emptiness Region n=1}) and (\ref{Eq: Equation of Emptiness Region n=2}), since $v$ diverges there. This leads to: 
\begin{align}
    \tau \approx \frac{x - 1}{v} + \frac{c_n}{v^{2n+3}},
    \label{Eq: polynomial equation around tau = 0}
\end{align}
where $c_n$ is a positive constant depending on $n$.
The linear term in $1/v$ shows that the empty region terminates at  $x =
1$. Unlike Eq. (\ref{Eq: cubic polynomial equation around x = 0}), the highest
power of the approximate polynomial equation is now depending on $n$ and this leads to the $n$-dependent power law of the emptiness boundary. One can again solve for the position of the local maximum and minimum of $\tau$ on $(x,\tau)$ to find 
\begin{align}
    1-x \propto |\tau|^{(2n+2)/(2n+3)}.
    \label{Eq: Power law near the end of emptiness}
\end{align}
Although Eq. (\ref{Eq: Power law near the end of emptiness}) is based on discrete value of $n$, we expect that the solution is deformed continuously  with the polytropic index $\gamma$  leading to Eq. (\ref{Eq: power law around tau = 0 in gamma notation}). The power law exponent is linear when $\gamma \rightarrow 1$  and square root when $\gamma \rightarrow \infty$.

Since the analytic instanton solutions are available for non-negative integer $n$, we are only able to write down the analytic expressions of the whole emptiness boundary for these $n$. Here we show the results for the emptiness boundary for $n = 0, 1$ and $2$
\begin{align}
&x = (1-\tau^{2/3})^{3/2},\label{Eq: Emptiness Boundary n=0}\\
&x = \Big[ 1-(2\tau)^{2/5} \Big]^{3/2} \Big[ 1+\frac{3}{2}(2\tau)^{2/5} \Big], \label{Eq: Emptiness Boundary n = 1}\\
&x = \Big[1-\left(\frac{8\tau}{3}\right)^{2/7}\Big]^{3/2} \Big[ 1 + \frac{3}{2}\left(\frac{8\tau}{3}\right)^{2/7}+\frac{15}{8}\left(\frac{8\tau}{3}\right)^{4/7} \Big]. \label{Eq: Emptiness Boundary n = 2}
\end{align} 
These results are summarized in Fig. \ref{fig: Emptiness
  Boundary}. Equation~(\ref{Eq: Emptiness Boundary n=0}) is  called the
astroid curve: $x^{2/3} + \tau^{2/3}=1$. For a general $n$ we have a  family of
astroid-like curves, like  Eqs.~(\ref{Eq: Emptiness Boundary n = 1}), (\ref{Eq: Emptiness Boundary n = 2}).

Having established the shape of the empty region, we now look at the behavior of the density profile immediately outside  of 
the empty region. The density grows from zero as a positive power of the distance from the boundary when $\tau \neq 0$ and diverges at $\tau = 0$,  Fig.~\ref{fig: Density}. The emptiness boundary is the branch point of the Riemann invariants $\lambda$ and $\overline{\lambda}$. The exponent of the power law can be determined from the degree of this branch point. Denoting the boundary of the empty region as $x_0=x_0(\tau)$, Eqs.~(\ref{Eq: Emptiness Boundary n=0})--(\ref{Eq: Emptiness Boundary n = 2}), one finds
\begin{align}
    \rho \propto (x-x_0)^{1/(\gamma - 1)},\qquad  \text{$x > x_0$}, 
                                         		\label{Eq: density around x = x0}
\end{align}
for $\tau\neq 0$ and 
\begin{align}
    \rho \propto (x-1)^{-2/(\gamma + 1)},\qquad  \text{$x > 1$}, 
    						\label{Eq: density around x = 1}
\end{align}
for $\tau=0$.

\section{Conclusions and Discussion}
\label{IV}

We have evaluated EFP for polytropic quantum liquids with $\gamma\geq 1$. Perhaps the most interesting application  
(besides previously well established fermion case  with $\gamma=3$) is the weakly interacting Bose gas. Within the Gross-Pitaevskii  approximation, its internal 
energy may be written as  \cite{pitaevskii2016bose} (in dimensionless form)
\begin{align}
V(\rho) = \frac{1}{2}\,\rho^2 -\rho + \frac{\xi^2}{R^2}\, \frac{(\partial_x \rho)^2}{8\rho} ,
\label{Eq: NLS equation of state}
\end{align}
where the last term constitutes the so-called quantum pressure. In the limit of the large emptiness, $R\gg \xi$, it is clearly sub-leading and the weakly interacting Bose gas is well approximated by the polytropic expression (\ref{Eq: Euclidean action V}) with $\gamma=2$.  It is interesting to note that it corresponds to the non-integer value $n=1/2$, which does not allow for an analytic solution of the hydrodynamic equations. Nevertheless we are able to deduce EFP through the analytic continuation procedure, resulting in
\begin{align}
-\ln {\cal P}_\mathrm{EFP}^\mathrm{Bose}(R) = \frac{\rho_0 R^2 }{\xi}\, \frac{16}{3\pi} + {\cal O}\left(\log(R/\xi)\right), 
\label{Eq: Bose EFP}
\end{align}
where $\xi\gg \rho_0^{-1}$ depends on the interaction strength. The logarithmic correction originates from the quantum pressure term in Eq.~(\ref{Eq: NLS equation of state}). This should be compared with the free fermion-RMT result \cite{des1973asymptotic,dyson1976fredholm} for $\gamma=3$:
\begin{align}
-\ln {\cal P}_\mathrm{EFP}^\mathrm{Fermi}(R) = \frac{\rho_0 R^2 }{\xi}\, \frac{\pi}{2} + \frac{1}{4} \log(R/\xi),
\label{Eq: Fermi EFP}
\end{align} 
where $\xi = (\pi\rho_0)^{-1}$. 

It is worth noticing that Eq.~(\ref{Eq: Bose EFP}) and (\ref{Eq: Fermi EFP}) are the two limiting cases of the integrable Lieb-Liniger model \cite{lieb1963exact}, for which the ground state wave function, $\Psi_\mathrm{GS}(x_1,x_2,...,x_N)$, is known explicitly through the Bethe Ansatz. The Bose and Fermi polytropic liquids represent its weak and strong interaction limits, correspondingly. While the correlation length, $\xi$, is known for any interaction strength, the coefficient multiplying 
$\rho_0 R^2/\xi$ so far is only available  in the two extreme limits. For intermediate interactions the equation of state of the Lieb-Liniger 
model is not polytropic. In fact, it interpolates between $\gamma=3$ at small density and $\gamma=2$ at large density. 
An appropriate solution of the hydrodynamic equations is not known for such equation of state. Nothing is known about the coefficient in front of the logarithmic correction, besides Tonks fermion limit, Eq.~(\ref{Eq: Fermi EFP}), either. In fact, it is not established if a term linear in $\rho_0 R$, is present in $-\log {\cal P}_\mathrm{EFP}$ at intermediate interactions (eg., the EFP in Calogero model  does contain such a linear term \cite{abanov2005hydrodynamics}).

There is an intriguing relation between imaginary time hydrodynamic for $n=0$, $\gamma=3$
and the density of states (DOS) in disordered superconductors with magnetic impurities 
\cite{larkin1965superconductor,feigel2000keldysh,savich2017quasiparticle}. The superconducting 
gap closes gradually with increasing concentration of the magnetic impurities. 
It turns out that the energy dependence of DOS is identical to spatial profile of the density  
in the top panel of Fig. \ref{fig: Density}, with magnetic impurity concentration playing the role of time. 
The details of this relation and its possible generalizations to $n>0$ are 
discussed in Appendix \ref{Appendix: Emptiness Formation and Disordered
  Superconductor}.

Another possibility to observe the spacetime shape of the emptiness arises
naturally in the context of the well known mapping of (1+1)D quantum field
theory onto 2D statistical models. The instanton solution of the former corresponds 
to stationary configuration dominating some
statistical mechanics models such as random tilings and  crystal surfaces, 
subject to proper boundary conditions \cite{kenyon2007limit,Jean-MarieReview2021}. 
In fact, there is a one to one
correspondence between random tilings and crystal surface heights on the  one hand  and  world
lines of free fermions in imaginary time on the other hand,  which was established via transfer matrix
representation of the partition function of these statistical models. It is
interesting to find statistical models with coarse grained properties  described by
polytropic equation of state  with $\gamma\neq 3$.

To conclude, we have developed a systematic way to construct analytic emptiness formation 
solution of hydrodynamic equations for polytropic liquids with the polytropic index $\gamma = (2n+3)/(2n+1)$, 
where $n$ is a non-negative integer. We evaluate the EFP and analytically continue the result 
to access EFP in polytropic liquid with an arbitrary $\gamma \geq 1$.  In particular, it yields a novel result for weakly interacting bosons, which may be experimentally verified in cold atom systems.

\section{Acknowledgments} 

We are indebted to A. Abanov, A. Gorsky and B. Meerson for valuable discussions. H-C.Y and AK were supported by NSF grant DMR- 2037654.

\appendix
\section{Hodograph Transformation}
\label{Appendix: Hodograph}
We want to go from spacetime $(x,\tau)$ to Riemann invariants $(\lambda, \overline{\lambda})$ as new coordinates. Following chain rules, partial derivatives with respect to $\lambda$ and $\overline{\lambda}$ can be expressed in terms of $x$ and $\tau$
\begin{align}
\begin{pmatrix}
\partial_\lambda \\
\partial_{\overline{\lambda}}
\end{pmatrix}
=
\begin{pmatrix}
\partial_{\lambda} x & \partial_{\lambda} \tau \\
\partial_{\overline{\lambda}} x & \partial_{\overline{\lambda}} \tau
\end{pmatrix}
\begin{pmatrix}
\partial_x \\
\partial_\tau
\end{pmatrix}.
\end{align}
Let's invert this equation,
\begin{align}
\begin{pmatrix}
\partial_x \\
\partial_\tau
\end{pmatrix}
&=
\frac{1}{J}
\begin{pmatrix}
\partial_{\overline{\lambda}} \tau & -\partial_{\lambda} \tau \\
-\partial_{\overline{\lambda}} x & \partial_{\lambda} x
\end{pmatrix}
\begin{pmatrix}
\partial_\lambda \\
\partial_{\overline{\lambda}}
\end{pmatrix}, \label{Eq: hodograph tranformation} \\
J &= \partial_{\lambda} x \partial_{\overline{\lambda}} \tau - \partial_{\lambda} \tau \partial_{\overline{\lambda}} x,
\end{align}
where  the Jacobian $J$ and is assumed to be nonzero. This is called Hodograph transformation. With Eq. (\ref{Eq: hodograph tranformation}), Eq. (\ref{Eq: Riemann Invariants-1}) and (\ref{Eq: Riemann Invariants-2}) become
\begin{align}
&\partial_{\overline{\lambda}} x - w \partial_{\overline{\lambda}}\tau = 0,\\
&\partial_{\lambda} x - \overline{w} \partial_{\lambda}\tau = 0.
\end{align}
These are the Eq. (\ref{Eq: Hodograph-1}) and (\ref{Eq: Hodograph-2}).

\section{Relation between EFP and the density asymptotic}
\label{Appendix: Analytic formula for EFP}

The following derivation is based on unpublished notes of A. Abanov. The exponent of EFP is related to the instanton action
\begin{align}
-\ln {\cal P}_\mathrm{EFP} = 2\text{Im} \left[S_{\mathrm{inst}}-S_0\right].
\end{align}
The factor of two in front the action allows to extend imaginary time integration to run from $-\infty$ to $+\infty$. The solutions at $\tau > 0$ is determined from time reversal symmetry: $\rho(x,\tau) = \rho(x, -\tau)$ and $j(x,\tau) = -j(x,-\tau)$.  The action can be formulated as a space-imaginary time integral
\begin{align}
2i(S_{\mathrm{inst}}-S_0) = \int_{-\infty}^{\infty}\int_{-\infty}^{\infty} dx d\tau \left[ \frac{mj^2}{2\rho} + V(\rho) - V(\rho_0) \right],
\end{align}
where we consider the polytropic equation of state $V(\rho)$ from Eq. (\ref{Eq: Equation of state}). By performing the variation of action with respect to $\rho$, $j$ and $\rho_0$, one gets
\begin{align}
&2i \delta ( S_{\mathrm{inst}}- S_0) = \nonumber\\
&\int_{-\infty}^{\infty}\int_{-\infty}^{\infty} dx d\tau \left\{ \frac{mj}{\rho} \delta j + \left [ -\frac{mj^2}{2\rho^2} + \partial_\rho V(\rho) \right] \delta \rho + \Big[ \partial_{\rho_0} V(\rho) - \partial_{\rho_0} V(\rho_0) \Big] \delta \rho_0 \right\},
\end{align}
However $\rho$, $j$ and $\rho_0$ are not independent of each other. They are constrained by the continuity relation. Introducing the displacement field $u$,  as 
\begin{align}
    \rho = \rho_0 + \partial_x u,\qquad \qquad 
    j = - \partial_\tau u,
\end{align}
allows to automatically resolve the continuity constraint. The variation of action now takes the form
\begin{align}
    2i \delta (S_{\mathrm{inst}}-S_0)
    &= \int_{-\infty}^{\infty}\int_{-\infty}^{\infty} dx d\tau \left\{ -\frac{mj}{\rho}\partial_\tau \delta u + \left[ -\frac{mj^2}{2\rho^2} + \partial_\rho V(\rho) \right] \partial_x \delta u \right. \nonumber \\ 
    & + \left. \left[ -\frac{mj^2}{2\rho^2} +  \partial_\rho V(\rho) +  \partial_{\rho_0} V(\rho) -  \partial_{\rho_0} V(\rho_0)  \right] \delta \rho_0  \right\}.
\end{align}
After integration by parts, one arrives  
\begin{align}
    2i \delta (S_{\mathrm{inst}}-S_0)
    &=\int_{-\infty}^{\infty}\int_{-\infty}^{\infty} dx d\tau \left\{ \left[ \partial_\tau \left(\frac{mj}{\rho}\right) - \partial_x \left( - \frac{mj^2}{2\rho^2} + \partial_\rho V(\rho) \right)  \right]\delta u \right. \nonumber \\ 
    & + \left. \left[ -\frac{mj^2}{2\rho^2} +  \partial_\rho V(\rho) +  \partial_{\rho_0} V(\rho) -  \partial_{\rho_0} V(\rho_0)  \right] \delta \rho_0  \right\},
\end{align}
where the first square bracket is zero on the equation of motion (\ref{Eq: Hydrodynamic Eq-2}), with the velocity field $v = j/\rho$. The boundary term is discarded since it vanishes at infinity. Only the second square bracket  contributes to the variation of the action. Then one substitutes the equation of state (\ref{Eq: Equation of state}) into integral and notices the sound velocity depending on $\rho_0$: $v_s \propto \rho_0^{(\gamma-1)/2}$. By rescaling the variables to dimensionless coordinates and fields: $x \rightarrow Rx,\ \tau \rightarrow R \xi m \tau,\ \rho \rightarrow \rho_0 \rho$ and $j \rightarrow (\rho_0/m\xi)j$, the action taken derivative with respect to average density is
\begin{align}
    2i\partial_{\rho_0} (S_{\mathrm{inst}}-S_0)
    = \frac{R^2}{\xi} \int_{-\infty}^{\infty}\int_{-\infty}^{\infty} dx d\tau \left[ -\frac{v^2}{2} + \frac{\rho^{\gamma-1}-1}{\gamma - 1} -(\rho - 1)  \right],
\end{align} The goal is to massage this space-imaginary time integral into a integration on the boundary at infinity. The first step is integration by parts in $x$
\begin{align}
    -\frac{v^2}{2} = x v\partial_x v - \partial_x \left(\frac{x v^2}{2}\right).
\end{align}
Using  equation of motion (\ref{Eq: Hydrodynamic Eq-2}) for $v$,
\begin{align}
    -\frac{v^2}{2} = -\partial_\tau \left(xv\right) + x \rho^{\gamma-2}
  \partial_x \rho - \partial_x \left(\frac{x v^2}{2}\right)\, ,
\end{align}
the integral becomes
\begin{align}
    &2i\partial_{\rho_0} (S_{\mathrm{inst}}-S_0)
    = \nonumber\\ 
    &\frac{R^2}{\xi} \int_{-\infty}^{\infty}  \int_{-\infty}^{\infty} dx d\tau \left[ -\partial_\tau (xv) + x \rho^{\gamma-2} \partial_x \rho - \partial_x \left(\frac{x v^2}{2} \right) + \frac{\rho^{\gamma-1}-1}{\gamma - 1} -(\rho - 1)  \right].
\end{align}
The terms with density can be absorbed into a total spatial derivative
\begin{align}
    &x \rho^{\gamma-2} \partial_x \rho + \frac{\rho^{\gamma-1}-1}{\gamma - 1} -(\rho - 1) = \partial_x \left(x\frac{\rho^{\gamma-1}-1}{\gamma-1} - u \right),
\end{align}
where $u$ is the dimensionless displacement field, $\partial_x u = (\rho -1)$. Now, one can apply Stokes' theorem
\begin{align}
    2i\partial_{\rho_0} (S_{\mathrm{inst}}-S_0)
    =\frac{R^2}{\xi} \oint \left[ (xv) dx + \left(x\frac{\rho^{\gamma-1}-1}{\gamma-1} - u -\frac{x v^2}{2} \right) d\tau \right].
\end{align}
On the boundary at the infinity, $v \approx -\partial_\tau u$ and $(\rho^{\gamma-1} -1) \approx (\gamma -1)(\rho -1) = (\gamma -1) \partial_x u$. The $v^2$ term decays too fast to give contribution in the integral. The boundary integral is further simplified as
\begin{align}
    2i\partial_{\rho_0} (S_{\mathrm{inst}}-S_0)
    = \frac{R^2}{\xi} \oint \Big[ (-x\partial_\tau u) dx + (x\partial_x u - u) d\tau \Big].
\end{align}
By defining complex variable, $z = x + i \tau$, the integral is performed on the complex $z$-plane
\begin{align}
    2i\partial_{\rho_0} (S_{\mathrm{inst}}-S_0)
    = \frac{R^2}{\xi} \oint \left[ \frac{-i}{2} \Big( \left(z+\overline{z}\right)\partial_z u - u \Big) dz + c.c. \right].
\end{align}
In general, the asymptotic of density at infinity is given by
\begin{align}
    \rho \approx 1 + \frac{\alpha}{2} \left(\frac{1}{z^2}+\frac{1}{\overline{z}^2} \right),
\end{align}
where $\alpha$ is some constant depending on polytropic index $\gamma$. At $\tau = 0$, it becomes the Eq. (\ref{Eq: asymptotics of density at x goes to infty}). The corresponding displacement field $u$ at infinity is given by
\begin{align}
    u \approx -\frac{\alpha}{2} \left(\frac{1}{z}+\frac{1}{\overline{z}} \right).
\end{align}
Substituting asymptotic of $u$ into the integral and performing integration in polar coordinates: $z = r \exp(i\theta)$ and $\overline{z} = r \exp(-i\theta)$, one arrives at Eq.~(\ref{Eq: derivative lnP and alpha}): 
\begin{align}
  i\partial_{\rho_0} (S_{\mathrm{inst}}-S_0)
    = \frac{R^2}{\xi}\, \frac{\alpha}{2} \int\limits_0^{2\pi}  (1 + \cos 2\theta ) d\theta
    = \frac{\pi R^2 \alpha}{\xi}.
\end{align}

\section{Emptiness Formation and Superconductor with magnetic impurity}
\label{Appendix: Emptiness Formation and Disordered Superconductor}

The action of the disordered superconductor with broken time reversal invariance, eg. by magnetic impurities, can be represented by the non-linear sigma model \cite{savich2017quasiparticle} as
\begin{align}
iS[Q] \propto \mathrm{Tr} \left\{-\frac{\eta}{2} [\sigma_z, Q]^2 + 4i\epsilon(\sigma_z Q) + 4i\Delta (i\sigma_y Q) \right\},
\label{Eq: Superconductor Action}
\end{align}
where  $\sigma_{x,y,z}$  are the Pauli matrices in the Nambu space, $\eta$ is the magnetic impurities concentration, $\epsilon$ is the energy and $\Delta$ is the superconducting order parameter. The gradient terms is neglected in the action under the assumption that vector potential varies slowly on the scale of superconducting correlation length. The Green function, $Q$, is the Nambu (and Keldysh) matrix constrained by $Q^2 = 1$.  Its retarded component may be parametrized in the  Nambu space as
\begin{align}
Q = \begin{pmatrix}
\cosh\theta & \sinh \theta \\
-\sinh\theta & -\cosh\theta
\end{pmatrix},
\label{Eq: Q matrix}
\end{align}
where $\theta$ is the complex Nambu angle and the density of states (DOS) is given by $\rho(\epsilon,\eta)$ $=\text{Re}[\cosh\theta]$. Performing variation over $\theta$, one obtains the saddle point equation
\begin{align}
\epsilon = \Delta \coth\theta - i\eta\cosh\theta,
\label{Eq. Superconductor Saddle}
\end{align}
which describes how DOS, $\text{Re}[\cosh\theta]$, changes with the magnetic impurity strength $\eta$.

Back to the hydrodynamics at $n = 0, \gamma = 3$, Riemann invariants, $\lambda$ and $\overline{\lambda}$, are decoupled. Solution of Eq.~(\ref{Eq: electrostatic-like V}) with the  boundary condition that density is zero within $|x| < R$ at $\tau = 0$ is
\begin{align}
x - \lambda \tau = R\, \frac{\lambda}{\sqrt{\lambda^2+1}},
\end{align}
where $R$ is the size of emptiness.  Changing  variables as $\lambda = i\cosh\theta$ this solution can be formulated as
\begin{align}
x =  R \coth\theta + i \tau\cosh\theta.
\label{Eq: Hydro n = 0}
\end{align}
One can establish correspondence between Eq. (\ref{Eq. Superconductor Saddle}) and (\ref{Eq: Hydro n = 0}) by identifying coordinate $x$ as energy $\epsilon$, emptiness size $R$ as superconducting order parameter $\Delta$ and imaginary time $-\tau$ as magnetic impurities concentration $\eta$. This indicates that DOS of a disordered superconductor is equivalent to the density of the 1D liquid with $n = 0$, forming the emptiness. The observation moment $\tau=0$ corresponds to the BCS time-reversal invariant case without magnetic impurities, where the gap is given by $\Delta$. Away from this limit the gap is suppressed  by the magnetic impurities until one reaches a gapless state at some critical $\eta_c$. The shape of the gap on the $(\epsilon,\eta)$ plane is given by the astroid $(\epsilon/\Delta)^{2/3}+(\eta/\eta_c)^{2/3}=1$ \cite{larkin1965superconductor}.

One may wonder if there are non-linear sigma model representations of the polytropic liquids with $n>0$, such that their 
densities coincide with DOS of corresponding superconductors. To this end we parametrize the velocity field as  
\begin{align}
    v = i(2n+1)\cosh\theta.
\end{align}
The equations Eq. (\ref{Eq: Equation of Emptiness Region n=1}) and (\ref{Eq: Equation of Emptiness Region n=2}) for $n = 1, 2$  become
\begin{align}
&x = R \left(\frac{3}{2}\coth \theta -\frac{1}{2}\coth^3 \theta \right) + 3i \tau \cosh \theta,
\label{Eq: Emptiness Region for n = 1 theta form}\\
&x = R \left(\frac{15}{8}\coth\theta -\frac{5}{4}\coth^3\theta + \frac{3}{8}\coth^5\theta \right) + 5i\tau \cosh\theta.
\label{Eq: Emptiness Region for n = 2 theta form}
\end{align}
One can construct the corresponding non-linear sigma models, leading to such saddle point equations (with the identification $x\to \epsilon$, $R\to\Delta$ and $\tau \to \eta$).  

For $n = 1$,
\begin{align}
iS[Q] \propto \mathrm{Tr}& \left\{ -\frac{3}{16}\eta [\sigma_z, Q]^4 +i\epsilon \left( \frac{1}{3} (\sigma_z Q )^3 - 3(\sigma_z Q) \right) + i\Delta \left( \frac{1}{3}(i\sigma_y Q)^3-3(i\sigma_y Q) \right)   \right\} .
\end{align}
And for $n = 2$,
\begin{align}
iS[Q] \propto \mathrm{Tr} & \left\{ -\frac{5}{16}\eta [\sigma_z, Q]^6 \right. + i \epsilon \left( \frac{3}{10}(\sigma_z Q)^5 -\frac{5}{2}(\sigma_z Q)^3 + 15(\sigma_z Q) \right) \nonumber\\
&+ \left. i\Delta \left( \frac{3}{10}(i\sigma_y Q)^5 - \frac{5}{2}(i\sigma_y Q)^3 + 15(i\sigma_y Q) \right)  \right\} .
\end{align}
However, the underlying microscopic models for these non-linear  sigma models are yet to be identified.

\bibliography{polytropic.bib}

\begin{thebibliography}{10}
\providecommand{\url}[1]{\texttt{#1}}
\providecommand{\urlprefix}{URL }
\expandafter\ifx\csname urlstyle\endcsname\relax
  \providecommand{\doi}[1]{doi:\discretionary{}{}{}#1}\else
  \providecommand{\doi}{doi:\discretionary{}{}{}\begingroup
  \urlstyle{rm}\Url}\fi
\providecommand{\eprint}[2][]{\url{#2}}

\bibitem{del2011long}
A.~del Campo,
\newblock \emph{Long-time behavior of many-particle quantum decay},
\newblock Phys. Rev. A \textbf{84}, 012113 (2011),
\newblock \doi{10.1103/PhysRevA.84.012113}.

\bibitem{pons2012fidelity}
M.~Pons, D.~Sokolovski and A.~del Campo,
\newblock \emph{Fidelity of fermionic-atom number states subjected to tunneling
  decay},
\newblock Phys. Rev. A \textbf{85}, 022107 (2012),
\newblock \doi{10.1103/PhysRevA.85.022107}.

\bibitem{del2016exact}
A.~del Campo,
\newblock \emph{Exact quantum decay of an interacting many-particle system: the
  calogero{\textendash}sutherland model},
\newblock New Journal of Physics \textbf{18}(1), 015014 (2016),
\newblock \doi{10.1088/1367-2630/18/1/015014}.

\bibitem{arzamasovs2019full}
M.~Arzamasovs and D.~M. Gangardt,
\newblock \emph{Full counting statistics and large deviations in a thermal 1d
  bose gas},
\newblock Phys. Rev. Lett. \textbf{122}, 120401 (2019),
\newblock \doi{10.1103/PhysRevLett.122.120401}.

\bibitem{esteve2006observations}
J.~Esteve, J.-B. Trebbia, T.~Schumm, A.~Aspect, C.~I. Westbrook and
  I.~Bouchoule,
\newblock \emph{Observations of density fluctuations in an elongated bose gas:
  Ideal gas and quasicondensate regimes},
\newblock Phys. Rev. Lett. \textbf{96}, 130403 (2006),
\newblock \doi{10.1103/PhysRevLett.96.130403}.

\bibitem{armijo2010probing}
J.~Armijo, T.~Jacqmin, K.~V. Kheruntsyan and I.~Bouchoule,
\newblock \emph{Probing three-body correlations in a quantum gas using the
  measurement of the third moment of density fluctuations},
\newblock Phys. Rev. Lett. \textbf{105}, 230402 (2010),
\newblock \doi{10.1103/PhysRevLett.105.230402}.

\bibitem{jacqmin2011sub}
T.~Jacqmin, J.~Armijo, T.~Berrada, K.~V. Kheruntsyan and I.~Bouchoule,
\newblock \emph{Sub-poissonian fluctuations in a 1d bose gas: From the quantum
  quasicondensate to the strongly interacting regime},
\newblock Phys. Rev. Lett. \textbf{106}, 230405 (2011),
\newblock \doi{10.1103/PhysRevLett.106.230405}.

\bibitem{bethe1931theorie}
H.~Bethe,
\newblock \emph{Zur theorie der metalle},
\newblock Zeitschrift f{\"u}r Physik \textbf{71}(3-4), 205 (1931),
\newblock \doi{10.1007/BF01341708}.

\bibitem{korepin1994correlation}
V.~E. Korepin, A.~G. Izergin, F.~H. Essler and D.~B. Uglov,
\newblock \emph{Correlation function of the spin-12 xxx antiferromagnet},
\newblock Physics Letters A \textbf{190}(2), 182 (1994),
\newblock \doi{10.1016/0375-9601(94)90074-4}.

\bibitem{korepin1997quantum}
V.~E. Korepin, N.~M. Bogoliubov and A.~G. Izergin,
\newblock \emph{Quantum inverse scattering method and correlation functions},
  vol.~3,
\newblock Cambridge university press, Cambridge, UK (1997).

\bibitem{de2001six}
J.~de~Gier and V.~Korepin,
\newblock \emph{Six-vertex model with domain wall boundary conditions: variable
  inhomogeneities},
\newblock Journal of Physics A: Mathematical and General \textbf{34}(39), 8135
  (2001),
\newblock \doi{10.1088/0305-4470/34/39/312}.

\bibitem{boos2003emptiness}
H.~Boos, V.~Korepin and F.~Smirnov,
\newblock \emph{Emptiness formation probability and quantum
  knizhnik–zamolodchikov equation},
\newblock Nuclear Physics B \textbf{658}(3), 417 (2003),
\newblock \doi{10.1016/S0550-3213(03)00153-6}.

\bibitem{kleinert2009path}
H.~Kleinert,
\newblock \emph{Path integrals in quantum mechanics, statistics, polymer
  physics, and financial markets},
\newblock World scientific, Singapore (2009).

\bibitem{mehta2004random}
M.~L. Mehta,
\newblock \emph{Random matrices}, vol. 142,
\newblock Elsevier, Amsterdam (2004).

\bibitem{dyson1962statisticalII}
F.~J. Dyson,
\newblock \emph{Statistical theory of the energy levels of complex systems.
  ii},
\newblock Journal of Mathematical Physics \textbf{3}(1), 157 (1962),
\newblock \doi{10.1063/1.1703774}.

\bibitem{dyson1962statisticalIII}
F.~J. Dyson,
\newblock \emph{Statistical theory of the energy levels of complex systems.
  iii},
\newblock Journal of Mathematical Physics \textbf{3}(1), 166 (1962),
\newblock \doi{10.1063/1.1703775}.

\bibitem{shiroishi2001emptiness}
M.~Shiroishi, M.~Takahashi and Y.~Nishiyama,
\newblock \emph{Emptiness formation probability for the one-dimensional
  isotropic xy model},
\newblock Journal of the Physical Society of Japan \textbf{70}(12), 3535
  (2001),
\newblock \doi{10.1143/JPSJ.70.3535}.

\bibitem{kitanine2000correlation}
N.~Kitanine, J.~Maillet and V.~Terras,
\newblock \emph{Correlation functions of the xxz heisenberg spin-12 chain in a
  magnetic field},
\newblock Nuclear Physics B \textbf{567}(3), 554 (2000),
\newblock \doi{10.1016/S0550-3213(99)00619-7}.

\bibitem{kitanine2002emptiness}
N.~Kitanine, J.~Maillet, N.~Slavnov and V.~Terras,
\newblock \emph{Emptiness formation probability of the xxz spin-$1/2$
  heisenberg chain at $\delta$= $1/2$},
\newblock Journal of Physics A: Mathematical and General \textbf{35}(27), L385
  (2002),
\newblock \doi{10.1088/0305-4470/35/27/102}.

\bibitem{Kitanine_2002}
N.~Kitanine, J.~M. Maillet, N.~A. Slavnov and V.~Terras,
\newblock \emph{Large distance asymptotic behaviour of the emptiness formation
  probability of {theXXZspin}-1/2 heisenberg chain},
\newblock Journal of Physics A: Mathematical and General \textbf{35}(49), L753
  (2002),
\newblock \doi{10.1088/0305-4470/35/49/102}.

\bibitem{bastianello2018exact}
A.~Bastianello, L.~Piroli and P.~Calabrese,
\newblock \emph{Exact local correlations and full counting statistics for
  arbitrary states of the one-dimensional interacting bose gas},
\newblock Phys. Rev. Lett. \textbf{120}, 190601 (2018),
\newblock \doi{10.1103/PhysRevLett.120.190601}.

\bibitem{dykman1994large}
M.~I. Dykman, E.~Mori, J.~Ross and P.~Hunt,
\newblock \emph{Large fluctuations and optimal paths in chemical kinetics},
\newblock The Journal of chemical physics \textbf{100}(8), 5735 (1994),
\newblock \doi{10.1063/1.467139}.

\bibitem{elgart2006classification}
V.~Elgart and A.~Kamenev,
\newblock \emph{Classification of phase transitions in reaction-diffusion
  models},
\newblock Phys. Rev. E \textbf{74}, 041101 (2006),
\newblock \doi{10.1103/PhysRevE.74.041101}.

\bibitem{krapivsky2012void}
P.~L. Krapivsky, B.~Meerson and P.~V. Sasorov,
\newblock \emph{Void formation in diffusive lattice gases},
\newblock Journal of Statistical Mechanics: Theory and Experiment
  \textbf{2012}(12), P12014 (2012),
\newblock \doi{10.1088/1742-5468/2012/12/p12014}.

\bibitem{krajenbrink2021PRL}
A.~Krajenbrink and P.~Le~Doussal,
\newblock \emph{Inverse scattering of the zakharov-shabat system solves the
  weak noise theory of the kardar-parisi-zhang equation},
\newblock Phys. Rev. Lett. \textbf{127}, 064101 (2021),
\newblock \doi{10.1103/PhysRevLett.127.064101}.

\bibitem{krajenbrink2021inverse}
A.~Krajenbrink and P.~L. Doussal,
\newblock \emph{Inverse scattering solution of the weak noise theory of the
  kardar-parisi-zhang equation with flat and brownian initial conditions},
\newblock arXiv preprint arXiv:2107.13497  (2021).

\bibitem{chernykh2001large}
A.~I. Chernykh and M.~G. Stepanov,
\newblock \emph{Large negative velocity gradients in burgers turbulence},
\newblock Phys. Rev. E \textbf{64}, 026306 (2001),
\newblock \doi{10.1103/PhysRevE.64.026306}.

\bibitem{elgart2004rare}
V.~Elgart and A.~Kamenev,
\newblock \emph{Rare event statistics in reaction-diffusion systems},
\newblock Phys. Rev. E \textbf{70}, 041106 (2004),
\newblock \doi{10.1103/PhysRevE.70.041106}.

\bibitem{janas2016dynamical}
M.~Janas, A.~Kamenev and B.~Meerson,
\newblock \emph{Dynamical phase transition in large-deviation statistics of the
  kardar-parisi-zhang equation},
\newblock Phys. Rev. E \textbf{94}, 032133 (2016),
\newblock \doi{10.1103/PhysRevE.94.032133}.

\bibitem{abanov2005hydrodynamics}
A.~G. Abanov,
\newblock \emph{Hydrodynamics of correlated systems},
\newblock In {\'E}.~Br{\'e}zin, V.~Kazakov, D.~Serban, P.~Wiegmann and
  A.~Zabrodin, eds., \emph{Applications of Random Matrices in Physics}, pp.
  139--161. Springer Netherlands, Dordrecht,
\newblock ISBN 978-1-4020-4531-8 (2006).

\bibitem{yeh2020emptiness}
H.-C. Yeh and A.~Kamenev,
\newblock \emph{Emptiness formation probability in one-dimensional bose
  liquids},
\newblock Phys. Rev. A \textbf{101}, 023623 (2020),
\newblock \doi{10.1103/PhysRevA.101.023623}.

\bibitem{joseph2011observation}
J.~A. Joseph, J.~E. Thomas, M.~Kulkarni and A.~G. Abanov,
\newblock \emph{Observation of shock waves in a strongly interacting fermi
  gas},
\newblock Phys. Rev. Lett. \textbf{106}, 150401 (2011),
\newblock \doi{10.1103/PhysRevLett.106.150401}.

\bibitem{calogero1969ground}
F.~Calogero,
\newblock \emph{Ground state of a one-dimensional n-body system},
\newblock Journal of Mathematical Physics \textbf{10}(12), 2197 (1969),
\newblock \doi{10.1063/1.1664821}.

\bibitem{sutherland1971exact}
B.~Sutherland,
\newblock \emph{Exact results for a quantum many-body problem in one
  dimension},
\newblock Phys. Rev. A \textbf{4}, 2019 (1971),
\newblock \doi{10.1103/PhysRevA.4.2019}.

\bibitem{sutherland1972exact}
B.~Sutherland,
\newblock \emph{Exact results for a quantum many-body problem in one dimension.
  ii},
\newblock Phys. Rev. A \textbf{5}, 1372 (1972),
\newblock \doi{10.1103/PhysRevA.5.1372}.

\bibitem{franchini2010emptiness}
F.~Franchini and M.~Kulkarni,
\newblock \emph{Emptiness and depletion formation probability in spin models
  with inverse square interaction},
\newblock Nuclear Physics B \textbf{825}(3), 320 (2010),
\newblock \doi{https://doi.org/10.1016/j.nuclphysb.2009.09.005}.

\bibitem{andric1983large}
I.~Andrić, A.~Jevicki and H.~Levine,
\newblock \emph{On the large-n limit in symplectic matrix models},
\newblock Nuclear Physics B \textbf{215}(2), 307 (1983),
\newblock \doi{https://doi.org/10.1016/0550-3213(83)90218-3}.

\bibitem{polychronakos1995waves}
A.~P. Polychronakos,
\newblock \emph{Waves and solitons in the continuum limit of the
  calogero-sutherland model},
\newblock Phys. Rev. Lett. \textbf{74}, 5153 (1995),
\newblock \doi{10.1103/PhysRevLett.74.5153}.

\bibitem{stone2008classical}
M.~Stone, I.~Anduaga and L.~Xing,
\newblock \emph{The classical hydrodynamics of the
  calogero{\textendash}sutherland model},
\newblock Journal of Physics A: Mathematical and Theoretical \textbf{41}(27),
  275401 (2008),
\newblock \doi{10.1088/1751-8113/41/27/275401}.

\bibitem{kulkarni2009nonlinear}
M.~Kulkarni, F.~Franchini and A.~G. Abanov,
\newblock \emph{Nonlinear dynamics of spin and charge in spin-calogero model},
\newblock Phys. Rev. B \textbf{80}, 165105 (2009),
\newblock \doi{10.1103/PhysRevB.80.165105}.

\bibitem{olver1988hamiltonian}
P.~J. Olver and Y.~Nutku,
\newblock \emph{Hamiltonian structures for systems of hyperbolic conservation
  laws},
\newblock Journal of mathematical physics \textbf{29}(7), 1610 (1988),
\newblock \doi{10.1063/1.527909}.

\bibitem{brunelli1997lax}
J.~Brunelli and A.~Das,
\newblock \emph{A lax description for polytropic gas dynamics},
\newblock Physics Letters A \textbf{235}(6), 597 (1997),
\newblock \doi{10.1016/S0375-9601(97)00708-1}.

\bibitem{brunelli2004integrable}
J.~Brunelli and A.~Das,
\newblock \emph{On an integrable hierarchy derived from the isentropic gas
  dynamics},
\newblock Journal of mathematical physics \textbf{45}(7), 2633 (2004),
\newblock \doi{10.1063/1.1756699}.

\bibitem{whitham2011linear}
G.~B. Whitham,
\newblock \emph{Linear and nonlinear waves}, vol.~42,
\newblock John Wiley \& Sons (2011).

\bibitem{sommerfeld1949partial}
A.~Sommerfeld,
\newblock \emph{Partial differential equations in physics},
\newblock Academic press (1949).

\bibitem{kamchatnov2000nonlinear}
A.~M. Kamchatnov,
\newblock \emph{Nonlinear periodic waves and their modulations: an introductory
  course},
\newblock World Scientific, Singapore (2000).

\bibitem{abanov2002probability}
A.~G. Abanov and V.~E. Korepin,
\newblock \emph{On the probability of ferromagnetic strings in
  antiferromagnetic spin chains},
\newblock Nuclear Physics B \textbf{647}(3), 565 (2002),
\newblock \doi{10.1016/S0550-3213(02)00899-4}.

\bibitem{abanov2003emptiness}
A.~G. Abanov and F.~Franchini,
\newblock \emph{Emptiness formation probability for the anisotropic xy spin
  chain in a magnetic field},
\newblock Physics Letters A \textbf{316}(5), 342 (2003),
\newblock \doi{10.1016/j.physleta.2003.07.009}.

\bibitem{franchini2005asymptotics}
F.~Franchini and A.~G. Abanov,
\newblock \emph{Asymptotics of toeplitz determinants and the emptiness
  formation probability for the {XY} spin chain},
\newblock Journal of Physics A: Mathematical and General \textbf{38}(23), 5069
  (2005),
\newblock \doi{10.1088/0305-4470/38/23/002}.

\bibitem{landau1987fluid}
L.~D. Landau and E.~M. Lifshitz,
\newblock \emph{Fluid mechanics},
\newblock Fluid Mechanics. Second Edition. 1987. Pergamon, Oxford  (1987).

\bibitem{trubnikov1987unstable}
B.~Trubnikov and S.~Zhdanov,
\newblock \emph{Unstable quasi-gaseous media},
\newblock Physics Reports \textbf{155}(3), 137 (1987),
\newblock \doi{10.1016/0370-1573(87)90123-2}.

\bibitem{matytsin1994large}
A.~Matytsin,
\newblock \emph{On the large-n limit of the itzykson-zuber integral},
\newblock Nuclear Physics B \textbf{411}(2), 805 (1994),
\newblock \doi{10.1016/0550-3213(94)90471-5}.

\bibitem{vilenkin2014extreme}
A.~Vilenkin, B.~Meerson and P.~V. Sasorov,
\newblock \emph{Extreme fluctuations of current in the symmetric simple
  exclusion process: a non-stationary setting},
\newblock Journal of Statistical Mechanics: Theory and Experiment
  \textbf{2014}(6), P06007 (2014),
\newblock \doi{10.1088/1742-5468/2014/06/p06007}.

\bibitem{meerson2014extreme}
B.~Meerson and P.~V. Sasorov,
\newblock \emph{Extreme current fluctuations in lattice gases: Beyond
  nonequilibrium steady states},
\newblock Phys. Rev. E \textbf{89}, 010101 (2014),
\newblock \doi{10.1103/PhysRevE.89.010101}.

\bibitem{Alex2016short-time}
A.~Kamenev, B.~Meerson and P.~V. Sasorov,
\newblock \emph{Short-time height distribution in the one-dimensional
  kardar-parisi-zhang equation: Starting from a parabola},
\newblock Phys. Rev. E \textbf{94}, 032108 (2016),
\newblock \doi{10.1103/PhysRevE.94.032108}.

\bibitem{smith2018landau}
N.~R. Smith, A.~Kamenev and B.~Meerson,
\newblock \emph{Landau theory of the short-time dynamical phase transitions of
  the kardar-parisi-zhang interface},
\newblock Phys. Rev. E \textbf{97}, 042130 (2018),
\newblock \doi{10.1103/PhysRevE.97.042130}.

\bibitem{Isoard2019wave}
M.~Isoard, A.~M. Kamchatnov and N.~Pavloff,
\newblock \emph{Wave breaking and formation of dispersive shock waves in a
  defocusing nonlinear optical material},
\newblock Phys. Rev. A \textbf{99}, 053819 (2019),
\newblock \doi{10.1103/PhysRevA.99.053819}.

\bibitem{kamchatnov2019landau}
A.~Kamchatnov,
\newblock \emph{Landau--khalatnikov problem in relativistic fluid dynamics},
\newblock Journal of Experimental and Theoretical Physics \textbf{129}(4), 607
  (2019),
\newblock \doi{10.1134/S1063776119100200}.

\bibitem{artin2015gamma}
E.~Artin,
\newblock \emph{The gamma function},
\newblock Courier Dover Publications, New York (2015).

\bibitem{arnol2003catastrophe}
V.~I. Arnol'd,
\newblock \emph{Catastrophe theory},
\newblock Springer Science \& Business Media, Berlin (2003).

\bibitem{berry1980iv}
M.~V. Berry and C.~Upstill,
\newblock \emph{Iv catastrophe optics: morphologies of caustics and their
  diffraction patterns},
\newblock In \emph{Progress in optics}, vol.~18, pp. 257--346. Elsevier,
  Amsterdam (1980).

\bibitem{bessis1984pole}
D.~Bessis and J.~Fournier,
\newblock \emph{{Pole condensation and the Riemann surface associated with a
  shock in Burgers' equation}},
\newblock {Journal de Physique Lettres} \textbf{45}(17), 833 (1984),
\newblock \doi{10.1051/jphyslet:019840045017083300}.

\bibitem{pitaevskii2016bose}
L.~Pitaevskii and S.~Stringari,
\newblock \emph{Bose-Einstein condensation and superfluidity}, vol. 164,
\newblock Oxford University Press, Oxford (2016).

\bibitem{des1973asymptotic}
J.~Des~Cloizeaux and M.~Mehta,
\newblock \emph{Asymptotic behavior of spacing distributions for the
  eigenvalues of random matrices},
\newblock Journal of Mathematical Physics \textbf{14}(11), 1648 (1973),
\newblock \doi{10.1063/1.1666239}.

\bibitem{dyson1976fredholm}
F.~J. Dyson,
\newblock \emph{{Fredholm determinants and inverse scattering problems}},
\newblock Communications in Mathematical Physics \textbf{47}(2), 171  (1976),
\newblock \doi{cmp/1103899727}.

\bibitem{lieb1963exact}
E.~H. Lieb and W.~Liniger,
\newblock \emph{Exact analysis of an interacting bose gas. i. the general
  solution and the ground state},
\newblock Phys. Rev. \textbf{130}, 1605 (1963),
\newblock \doi{10.1103/PhysRev.130.1605}.

\bibitem{larkin1965superconductor}
A.~Larkin,
\newblock \emph{Superconductor of small dimensions in a strong magnetic field},
\newblock Sov. Phys. JETP \textbf{21}, 153 (1965).

\bibitem{feigel2000keldysh}
M.~V. Feigel'man, A.~I. Larkin and M.~A. Skvortsov,
\newblock \emph{Keldysh action for disordered superconductors},
\newblock Phys. Rev. B \textbf{61}, 12361 (2000),
\newblock \doi{10.1103/PhysRevB.61.12361}.

\bibitem{savich2017quasiparticle}
Y.~Savich, L.~Glazman and A.~Kamenev,
\newblock \emph{Quasiparticle relaxation in superconducting nanostructures},
\newblock Phys. Rev. B \textbf{96}, 104510 (2017),
\newblock \doi{10.1103/PhysRevB.96.104510}.

\bibitem{kenyon2007limit}
R.~Kenyon and A.~Okounkov,
\newblock \emph{{Limit shapes and the complex Burgers equation}},
\newblock Acta Mathematica \textbf{199}(2), 263  (2007),
\newblock \doi{10.1007/s11511-007-0021-0}.

\bibitem{Jean-MarieReview2021}
J.-M. Stéphan,
\newblock \emph{{Extreme boundary conditions and random tilings}},
\newblock SciPost Phys. Lect. Notes p.~26 (2021),
\newblock \doi{10.21468/SciPostPhysLectNotes.26}.

\end{thebibliography}

\end{document}